\documentstyle[12pt,aaspp4]{article}
\def\plotone#1{\centerline{\epsfxsize=5.0in\epsfbox{#1}}}
\def\plotonesmall#1{\centerline{\epsfxsize=3.5in\epsfbox{#1}}}
\def\plotplate#1{\centerline{
 THIS FIGURE AVAILABLE at http://astro.berkeley.edu/dust/index.html}}
\def\BE{\begin{equation}}
\def\BEL#1{\begin{equation}\label{#1}}
\def\EE{\end{equation}}
\newcommand{\HI}{H\,{\scriptsize I}}
\newcommand{\HII}{H\,{\scriptsize II}}
\newcommand{\etal}{{\it et al.}}
\newcommand{\eg}{{\it e.g.}}

\newcommand{\cf}{{\it c.f.}}
\newcommand{\COBE}{{\it COBE}}
\newcommand{\IRAS}{{\it IRAS}}
\newcommand{\MAP}{{\it MAP}}
\newcommand{\Av}{{\rm A(V)}}
\newcommand{\Ebv}{E(B-V)}
\newcommand{\Rmap}{{\mathcal R}} 
\newcommand{\Rbar}{\bar{\Rmap}_i} 
\newcommand{\Wmap}{{\mathcal W}} 
\newcommand{\kapstar}{\kappa^\star} 

\newcommand{\Imeani}{\langle I_i \rangle}
\newcommand{\Tmean}{\langle T \rangle}
\newcommand{\Tmeanone}{\langle T_1 \rangle}
\newcommand{\Tmeantwo}{\langle T_2 \rangle}
\newcommand{\MAG}{{\rm ~mag}}
\newcommand{\Ang}{{\rm ~\AA}}

\newcommand{\degree}{^\circ}
\newcommand{\s}{{\rm ~s}}

\newcommand{\cm}{{\rm ~cm}}
\newcommand{\Jy}{{\rm ~Jy}}

\newcommand{\MJypSr}{{\rm ~MJy}/{\rm sr}}
\newcommand{\GHz}{{\rm ~GHz}}
\newcommand{\K}{{\rm ~K}}
\newcommand{\mK}{{\rm ~mK}}
\newcommand{\mm}{{\rm ~mm}}
\newbox\grsign \setbox\grsign=\hbox{$>$} \newdimen\grdimen \grdimen=\ht\grsign
\newbox\simlessbox \newbox\simgreatbox
\setbox\simgreatbox=\hbox{\raise.5ex\hbox{$>$}\llap
     {\lower.5ex\hbox{$\sim$}}}\ht1=\grdimen\dp1=0pt
\setbox\simlessbox=\hbox{\raise.5ex\hbox{$<$}\llap
     {\lower.5ex\hbox{$\sim$}}}\ht2=\grdimen\dp2=0pt
\def\simgt{\mathrel{\copy\simgreatbox}}\def\simlt{\mathrel{\copy\simlessbox}}

\begin{document}

\title{Extrapolation of Galactic Dust Emission at 100 Microns to CMBR
Frequencies Using FIRAS}

\author{Douglas P. Finkbeiner \& Marc Davis}
\affil{University of California at Berkeley, Departments of Physics and 
Astronomy, 601 Campbell Hall, Berkeley, CA 94720}
\authoremail{dfink@astro.berkeley.edu, marc@deep.berkeley.edu}
\and
\author{David J. Schlegel}
\affil{Princeton University, Department of Astrophysics,
Peyton Hall, Princeton, NJ 08544}
\authoremail{schlegel@astro.princeton.edu}


\begin{abstract}

We present predicted full-sky maps of submillimeter and microwave
emission from the diffuse interstellar dust in the Galaxy.  These maps
are extrapolated from the $100\micron$ emission and $100/240\micron$
flux ratio maps that Schlegel, Finkbeiner, \& Davis (1998; SFD98)
generated from \IRAS\ and \COBE/DIRBE data.  Results are presented for
a number of physically plausible emissivity models. 

The correlation of \COBE/FIRAS data with the simple SFD98 ($\nu^2$
emissivity power law) extrapolation is much tighter than with
other common dust templates such as \HI\ column density or $100\micron$
emission.  Despite the apparent success of the SFD98 extrapolation,
the assumed $\nu^2$ emissivity is inconsistent with the FIRAS data
below $800\GHz$.  Indeed, no power law emissivity function fits the
FIRAS data from $200 - 2100\GHz$.  In this paper we provide a
formalism for a multi-component model for the dust emission.  A
two-component model with a mixture of silicate and carbon-dominated
grains (motivated by \cite{pollack94}) provides a fit to an accuracy
of $\sim 15\%$ to all the FIRAS data over the entire high-latitude
sky.  Small systematic differences are found between the atomic and
molecular phases of the ISM.

\COBE/DMR has observed microwave emission that is correlated with thermal dust
emission.  However, this emission is higher than our model predicts by
factors of $1.2$, $2.4$ and $20$ at $90$, $53$ and $31\GHz$,
respectively.  This provides evidence that another emission
mechanism dominates dust emission at frequencies below $\sim 60\GHz$. 

Our predictions for the thermal (vibrational) emission from Galactic
dust at $\nu < 3000\GHz$ are available for general use.  These full-sky
predictions can be made at the DIRBE resolution of $40'$ or at the
higher resolution of $6{\farcm}1$ from the \cite{sfd98}
DIRBE-corrected \IRAS\ maps.
 
\end{abstract}


\section{INTRODUCTION}

The pioneering Infrared Astronomy Satellite (\IRAS) led to
the discovery of the ubiquitous infrared cirrus, whose thermal emission
is especially visible in the $100\micron$ band (\cite{low84}).
This cirrus, with a characteristic temperature of $\sim 20\K$,
arches across the sky in long filamentary chains and is present at all
Galactic latitudes.  However, \IRAS\ was optimized for the detection
of point sources, and its ability to map the diffuse cirrus was less
than optimal.  Because of calibration drifts and hysteresis effects,
the resulting \IRAS\ Sky Survey Atlas (ISSA: \cite{issa94}) images are
contaminated by significant striping and poor control of large scale gradients.

The Diffuse Infrared Background Experiment (DIRBE)
on the \COBE\ satellite is the perfect complement to \IRAS.
It has relatively low angular resolution ($0.7\degree$) but superbly
controlled zero-points and gains.  This has led to the
generation of a map of the far-infrared sky with unprecedented accuracy and
uniformity of coverage.  Schlegel, Finkbeiner, \& Davis (1998; hereafter
\cite{sfd98}) created a merged map of the \IRAS\ and DIRBE data with an
angular resolution of $6\arcmin$ and DIRBE-quality calibration.
Their full-sky map shows the pervasive
extent of the infrared cirrus and has proven successful for estimation of
extragalactic reddening.  But equally important will be the use of this
type of data for estimation of Galactic foreground for the coming generation
of CMBR experiments, including \MAP\ and Planck and a host of ground- and
balloon-based projects.
 
In this paper, we consider the use of the SFD98 dust map as a
predictor for microwave emission from Galactic dust.  The SFD98 map is
based solely upon $100-240\micron$ ($1250-3000\GHz$) emission.
Extrapolation to microwave frequencies is very sensitive to the
details of the composition and emissivity properties of the dust.  We
show that the $\nu^2$ emissivity assumed by SFD98 is inconsistent with
the $100-2100\GHz$ emission probed by the \COBE\ Far Infrared Absolute
Spectrophotometer (FIRAS).  We use these FIRAS data to constrain the
properties of the dust, and show that no power law emissivity model
can consistently explain the full spectral range of the dust emission.
However, we find excellent agreement with a two-component model whose
components we tentatively refer to as silicate and
carbon-dominated grains.  With this model for the dust emissivity function,
extrapolation of Galactic dust emission from $100\micron$ to lower
frequencies is based upon the filtered DIRBE $100/240\micron$ color
temperature.

In \S \ref{sec_data}, we discuss the \COBE\ data sets and the details
of comparisons using SFD98.
Section \ref{sec_1comp} explores a variety of one-component dust models,
demonstrating that a single power-law emissivity fails to explain the
data, as does a broadened temperature distribution.
Section \ref{sec_multicomp} explores a family of two-component dust models, 
in which energy balance and the temperature of the separate 
components are tightly coupled -- one of which achieves excellent
agreement with the FIRAS data.  Section \ref{sec_discussion}
discusses the robustness of this best model with respect to 
various ISM environments, and \S \ref{sec_dmr} compares
our predictions to (DMR) microwave observations,
demonstrating that the microwave emission may exceed the predictions of
any thermal (vibrational) emission mechanisms.  This is perhaps the
signature of spinning dust grains emitting electric dipole
radiation (\cite{draine98b}) or the signature of free-free emission.
Summary and conclusions are presented in \S \ref{sec_summary}.

\section{DATA SETS}
\label{sec_data}

The \COBE\ (COsmic Background Explorer) satellite consisted of three
instruments, DMR (Differential Microwave Radiometer), FIRAS (Far
Infrared Absolute Spectrophotometer), and DIRBE (Diffuse Infrared Background
Experiment).  In this paper we shall compare predictions of dust
emission based on DIRBE in the far-infrared with that observed by FIRAS at
lower frequencies.  In addition, we extend this correlation to still lower
frequencies ($31.5$, $53$, and $90 \GHz$) observed by DMR.  
Although the DMR fluctuations are dominated by intrinsic CMBR
anisotropy, a residual correlation with DIRBE is detectable even at
high latitudes.

\subsection{FIRAS Spectra}

The objective of the FIRAS instrument was to compare the cosmic
microwave background radiation (CMBR) to an accurate blackbody, and to
observe the dust and line emission from the Galaxy. It is a polarizing
Michelson interferometer (\cite{mather82}), operated differentially
with an internal reference blackbody and calibrated by an external
blackbody with an emissivity known to better than 1 part in $10^4$.
It covers the wavelength range from $0.1$ to $10 \mm$ ($30 - 3000 \GHz$)
in two spectral channels separated at $\sim0.5 \mm$ ($600 \GHz$).
The spectral resolution is $\sim 20\GHz$.
Although the design of the FIRAS experiment was optimized 
for its very successful measurement of the CMBR spectrum (\cite{fixsen96}),
the instrument also measured the spectrum
of the dust emission of our Galaxy (\eg, \cite{fixsen96}).
For the highest frequency channels, the Galactic signal dominates
all others.  

A flared horn antenna aligned with the \COBE\ spin
axis gives the FIRAS a $7\degree$ field of view.  The instrument
was cooled to $1.5 \K$ to reduce its thermal emission and enable the use
of sensitive bolometric detectors.  The FIRAS ceased to operate when
the \COBE\ supply of liquid helium was depleted on 21 September 1990, by
which time it had surveyed the sky 1.6 times.

We use the FIRAS Pass 4 Galactic dust spectra (hereafter FIRAS dust
spectra) from which CMBR, zodiacal light, and a FIRB model have been
subtracted (\cite{fixsen97}).
The data are presented as 213 spectral bins on the resolution 6
skycube map (6144 pixels on the full sky).\footnote{These data are
available on the World Wide Web at {http://www.gsfc.nasa.gov/astro/cobe/}.}
Several Galactic emission lines, such as C$^{+}$ ($157.7\micron$),
have been removed and replaced with interpolated values.
For our analyses, we have removed the troublesome frequency bins
listed in table \ref{table_badbins}, and recalibrated the entire FIRAS
data set down by 1\%.  Our analyses make use of 123
frequency bins at $100 < \nu < 2100 \GHz$ ($140\micron < \lambda < 3\mm$).
Note that data in the lowest two frequency bins are off the page in
some of the figures, but are used in the fits.  
Details of the frequency bin choice and recalibration can be found in
Appendix \ref{app_firas}.

\subsection{DMR Data}

DMR observed the sky at three frequencies, $31.5$, $53$, and $90
\GHz$, achieving the first detection of anisotropy in the CMBR
(\cite{smoot92}).  In this paper we use the 4-year DMR skymaps dated
18 April 1995, which have the monopole and dipole removed.  These maps
do not influence any of our model fits, but are compared with our
predictions in \S \ref{sec_dmr}.

Kogut \etal\ (1996) observed a correlation between Galactic dust and
the $31.5$ and $53 \GHz$ channels of DMR that is much greater than
that expected from \emph{any} models of thermal (vibrational) emission
by dust.  Alternative explanations such as spinning dust grains
(\cite{draine98b}) or spatially correlated free-free emission have
been proposed, but are not well constrained by existing data (\cf,
\cite{doc98a}).  We discuss this excess emission in \S \ref{sec_dmr}.

\subsection{DIRBE Data and SFD Dust Maps}
\subsubsection{SFD emission map}

\cite{sfd98} presented a full-sky $100\micron$ cirrus emission map
constructed from both the DIRBE and \IRAS/ISSA data sets.
The map is well calibrated, zodiacal light subtracted, Fourier destriped,
and point source subtracted, with a final resolution of $6.1\arcmin$.
Complete descriptions of these maps may be found in
\cite{sfd98}.\footnote{These data are publicly available via the
World Wide Web at {http://astro.berkeley.edu/dust}.}
For comparisons with FIRAS and DMR, the full resolution
of the \IRAS/DIRBE map is not required.  Instead, we use the
$0.7\degree$ DIRBE map with zodiacal light removed as described in SFD98,
with point sources included.  The DIRBE map offers a fair comparison
with the high-frequency FIRAS data, in which these sources contribute
to the measured flux.  The comparison is less appropriate in the low-frequency
FIRAS data, where typical FIR-luminous sources are expected to contribute
little to the measured millimeter flux.  However, the contribution from
stars and galaxies to the $100\micron$ flux is only $\sim 2\%$ of the
diffuse Galactic emission at high latitudes, and relatively less at
low latitudes.

\subsubsection{Ratio Map}

We also make use of a DIRBE $100\micron$/$240\micron$ color ratio
similar to that described in SFD98.
Because of the poor signal-to-noise ratio in the $240\micron$ map,
SFD98 employed a filtering algorithm to give the minimum variance estimate
of the dust temperature in each $1.3\degree$ Gaussian beam.
In each pixel this filter yields the weighted average of the measured flux
and a more robust estimator -- in the case of SFD98, the estimator is 
the $|b|>75\degree$ average flux.  The weights are chosen so that the
ratio of the filtered maps is the minimum variance estimate of the
true flux ratio.  The process gives the measured ratio in high S/N
pixels, but recovers the high latitude average ratio of 0.66 in the
limit of low S/N. 

The SFD98 algorithm has the unfortunate effect of suppressing
temperature variations at high latitude even when those variations are
measurable at a resolution of a few degrees.
In the current analysis, the S/N of FIRAS in a $7\degree$ beam is
sufficiently high that this non-local filtering algorithm causes
undesirable behavior in model fits.
When the DIRBE $100\micron$ and $240\micron$ maps
are smoothed to $7\degree$, structure in the ratio appears which is
not aligned with the imperfectly-subtracted zodiacal plane or other
potential artifacts in the maps.  Rather, the DIRBE $240\micron$ map
exhibits structure, even at very low levels, that is correlated with
the FIRAS maps at $240\micron$.  Therefore, it is presumed that this
structure is of extra-solar origin, and should not be discarded as it
was in the SFD98 analysis. 

In the current paper, we have constructed a new ratio map, $\Rmap$,
that retains more temperature information.  We use the same weight
function $\Wmap$ described in SFD98 (equations 8 and 9).  But rather
than forcing the map to a high-latitude average at low S/N, we force
it to the local $7\degree$ average.  High $S/N$ regions are little
changed from the previous $R$ map, but large-scale temperature
structures are now apparent at high Galactic latitudes that were
suppressed before. 
The temperature correction derived from this ratio map have a
$1.3\degree$ resolution in high $S/N$ regions, and is applied to the
full-resolution $100\mu$ map, not the smoothed $100\micron$ map.
This procedure correctly handles
the situation where a compact, high S/N source is located near a
diffuse background with a different color temperature. 
It should be noted here that the same 13 bright
sources listed in SFD98 Table 1 were removed from the DIRBE maps
before smoothing, to avoid halo artifacts in the $\Rmap$ map.  This
change in $\Rmap$ produces only very modest change in the SFD98
reddening predictions.  The largest change to predicted reddenings at
high latitude is of order $\Ebv = 0.01\MAG$.

\subsubsection{Cosmic IR Background Removal}

The ratio map and derived temperatures are moderately dependent upon
the uncertainties in cosmic infrared background (CIB).  The CIB
represents the extragalactic signal that is unresolved and isotropic
in either the DIRBE or FIRAS instruments.  This signal is presumably
from high-redshift ($z \ga 1$) dust-enshrouded galaxies, which are
only beginning to be resolved with ground-based submillimeter
observations (\cf, \cite{blain99}).  Detections of the CIB at $140$
and $240\micron$ were reported last year by \cite{sfd98} and
\cite{hauser98}, and upper limits were reported at $100\micron$.  A
more definitive analysis is in preparation by Finkbeiner, Schlegel, \&
Davis (1999).  We remove the CIB from the DIRBE maps in the same way
as \cite{sfd98} - as part of the zodiacal light model.  By using the
zero point of the Leiden-Dwingeloo \HI\ map, a model including CIB and
zodiacal light may be fit and removed.  This is an easier problem than
the separation of CIB from zodiacal light, which is unnecessary for
this paper.  One source of error in this could result if there is
significant dust emission correlated with H$\alpha$, and the
H$\alpha$/dust correlation has a different zero point than the \HI\
/dust correlation.  The sense of this would be to add a constant to
both $I_{100}$ and $I_{240}$, causing the derived temperature
distribution on the sky to broaden or narrow.  In other words, a poor
$\Rmap$ would produce FIRAS fit residuals that depend on temperature,
which would show up in figure \ref{fig_fitres_temp}.  Lack of a
temperature-dependent residual indicates that CIB removal errors have
had a negligible effect on our model.  The rest of our fit procedure -
using correlation slopes at each frequency - ignores an isotropic
component by construction, so we conclude that we are unaffected by
uncertainty in the CIB. 

\subsection{Comparing \COBE\ Data Sets}

\subsubsection{Beam Shapes}

Comparisons between DIRBE and FIRAS data are made at the FIRAS resolution.
The FIRAS beam has a frequency-dependent shape that is not well-measured.
The beam is approximately a $7\degree$-diameter tophat in the
highest-frequency channels, with power-law wings
(measured at $750\GHz$ from off-axis measurements of the moon;
see \S 7.9.4 of \cite{firas_supp}).
The beam shape is closer to Gaussian at lower frequencies, with exponential
wings from $5\degree$ to $15\degree$ from the beam center
(measured at $90\GHz$ in the lab; see \cite{fixsen94}).
Because the FIRAS scan strategy averages over 32 to 46 sec interferograms,
the beam is smeared by typically $2.3\degree$ in approximately lines
of constant ecliptic longitude.
The pixelization of the FIRAS data on $\sim 3\degree$ pixels introduces
another effective smoothing.
We approximately match the FIRAS beam by first convolving the DIRBE data
with a $7.0\degree$ circular tophat, then convolving with a $3.0\degree$
circular tophat, then smoothing by $2.3\degree$ in ecliptic longitude.
We ignore the non-Gaussian beam shapes of
the DIRBE instruments since they are sufficiently smaller.

We attempted to match the frequency-dependence of the FIRAS beam.
The signal in the Galactic plane is sufficiently strong
($\sim 100$ times larger than the median value at $500\GHz$)
that the exact sidelobe profile may be important.
At low frequencies, the sidelobes exceed $10^{-3}$ within $8\degree$
of the beam center.
However, since the profile has only been measured at two frequencies,
it is impossible to model the beam to high accuracy.
Therefore, the beam shape uncertainties introduce errors of up to $10\%$
within $7\degree$ of the plane.
Because we exclude the sky within $7\degree$ of the Galactic plane from
our analyses for other reasons (see \S \ref{sec_mask}),
we simply ignore the complication of frequency-dependence of the FIRAS beam.

\subsubsection{Spatial Mask}
\label{sec_mask}

Our analysis is limited to those parts of the sky where the far-infrared
emission is expected to be dominated by the diffuse interstellar medium.
We create a spatial mask that excludes the Galactic plane
below $|b| = 7\degree$, the Magellanic Clouds, and \HII\ regions in
Orion and Ophiuchus.
In such regions, the SFD temperature map is unreliable due to confusion limits.
These are also the regions where the FIRAS data suffer from poorly understood
sidelobe contamination.
We also mask $1.3\%$ of the sky where the FIRAS coverage is missing
or incomplete, and another $15\%$ where the FIRAS pixel weight is less
than 0.4 (the median value is 0.8). 
The final mask excludes $29\%$ of the sky from our analyses, and is shown
as the thin black outlines in Figs.\ \ref{plate_res_hi}, \ref{plate_res_100},
and \ref{plate_res_pred}.
This mask is used throughout this paper except for the comparison in
figure \ref{fig_3temp} in which the Galactic plane is included, and
for the comparisons with DMR shown in Tables \ref{table_dmr} and
\ref{table_dmr_x}.  For the DMR comparisons we apply the Goddard ``custom cut''
mask from the 4-year DMR data analysis which excludes 37\% of the sky 
(\cite{bennett96}).

\subsubsection{Simple Difference Spectra}
\label{sec_diff}

\begin{figure}
\plotone{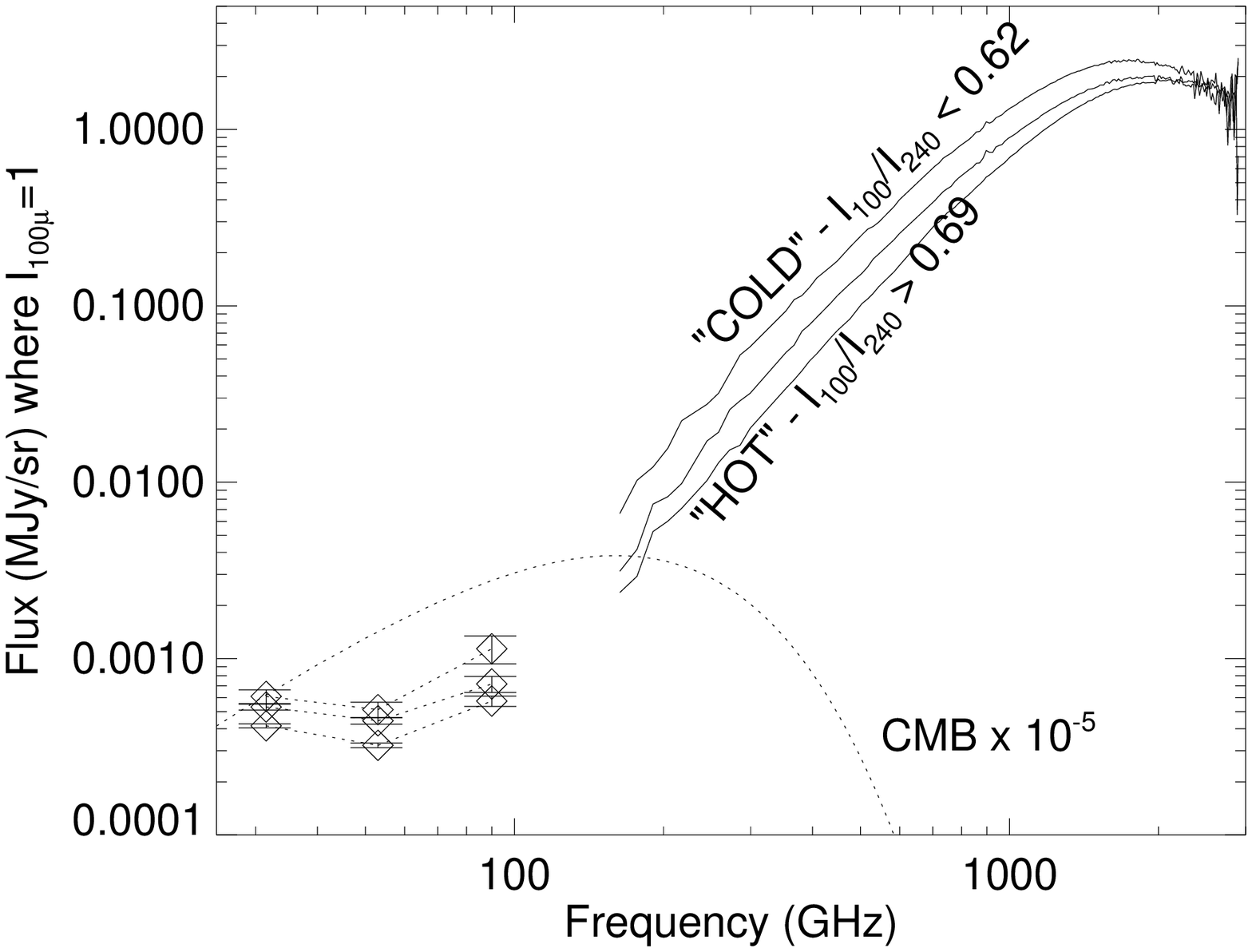}
\caption{Difference spectra from \COBE\ data, after CMBR monopole and
dipole removal.  The bright and faint regions of the sky are differenced
for each channel in the DMR (diamonds) and FIRAS (solid lines) data sets,
excluding the Galactic plane and Magellenic Clouds.
The sky is divided into cold, warm and hot zones based upon DIRBE
$I_{100}/I_{240}$ color ratios.
The differences in each zone are renormalized to a $100\micron$ flux of
$1.0\MJypSr$, which is a typical flux level for the high-latitude sky.
Note the factor of two difference between the cold and hot zones
at $\nu \simlt 700 \GHz$, relative to the $100\micron$ normalization.
For comparison, the dotted line represents $10^{-5}$ the level of the
CMBR spectrum.
}
\label{fig_3temp}
\end{figure}

For an overview of the three \COBE\ datasets, it is useful to plot the
DMR, FIRAS, and low-frequency DIRBE data on a single plot.
The DMR is a differential instrument, so the mean measurement
over the sky is zero in each channel.  In order to compare it to FIRAS
and DIRBE, we plot the difference between
``bright'' ($I_{900\GHz} > 3.0 \MJypSr$) and ``faint'' regions of the sky.  
This method has the additional
advantage of discarding any isotropic background of cosmic or
instrumental origin.  We have further divided the sky into cold,
warm, and hot zones according to the DIRBE $I_{100}/I_{240}$
color ratio.  The cold component ($I_{100}/I_{240} < 0.62$) comprises
$14\%$ of the sky, the hot component  ($I_{100}/I_{240} > 0.69$)
comprises $26\%$, and the warm component comprises $44\%$.
The remaining $16\%$ is masked, rejecting only bad or noisy FIRAS
pixels, but including signal in the Galactic plane for better S/N. 
This plot assumes a monotonic relationship between color ratio and
physical temperature, but requires no other knowledge of the dust spectra.
The difference spectra for these regions are shown in Figure
\ref{fig_3temp}.  All three curves are renormalized such that the DIRBE
$100\micron$ flux is $1\MJypSr$, a value typical for high-latitude dust.
The spectra qualitatively have the correct behavior, with the ``cold''
regions showing stronger emission at low frequencies relative to $100\micron$.

The FIRAS emission at low frequencies ($200 \la \nu \la 600\GHz$) scales as
$\sim \nu^{3.2}$.  Because the Planck function $B_\nu(T)$ asymptotes slowly
to $\nu^2$ on the Rayleigh-Jeans tail, the best fit emission spectrum is
$\sim \nu^{1.7} B_\nu(19\K)$ over this frequency range,
not $\sim \nu^{1.2} B_\nu(19\K)$. 
The temperature $19\K$ corresponds to the median dust temperature for
this fit to the emission spectrum.
These considerations indicate that the naive $\nu^2$ emissivity law assumed
in SFD98 is incorrect, a matter that will be explored extensively in this
paper.

Furthermore, Figure \ref{fig_3temp} demonstrates that the Galactic
emission detected by DMR is inconsistent with any power-law extrapolation
of the FIRAS data.  The DMR $31.5$ and $53 \GHz$ channels lie
well above the power law extrapolation of the FIRAS curves.
We address this problem in \S \ref{sec_dmr}.


\section{ONE-COMPONENT DUST MODELS}
\label{sec_1comp}

\subsection{Predicted Microwave Emission from SFD98}

A simple but naive prediction for submillimeter/microwave emission can be made
from our previous work.
\cite{sfd98} extensively studied the emission from dust in the regime
$100\micron < \lambda < 240\micron$.
Assuming a $\nu^2$ emissivity model, the temperature of the dust was
mapped with a resolution of $1.3\degree$ from
the DIRBE $100\micron$/$240\micron$ emission ratio, $\Rmap$.
The $100\micron$ emission of the dust was mapped with a resolution of
$6.1\arcmin$
by utilizing small-scale information from the \IRAS\ mission.
Emission at lower (millimeter/microwave) frequencies can be predicted by
extrapolating the $100\micron$ flux using this temperature fit.
For each line-of-sight in the maps, the emission at frequency $\nu$ can
be expressed as
\BEL{equ_extrap1}
    I_\nu = K^{-1}_{100}(\alpha,T) I_{100}
    {\nu^\alpha B_\nu(T) \over \nu_0^\alpha B_{\nu_0}(T) }
\EE
where $B_\nu(T)$ is the Planck function at temperature $T$,
$I_{100}$ is the DIRBE-calibrated $100\micron$ map, $K_{100}(\alpha,T)$
is the color correction factor for the DIRBE $100\micron$ filter when
observing a $\nu^\alpha B_\nu(T)$ spectrum, and $\nu_0=3000\GHz$ is
the reference frequency corresponding to $100\micron$.
Our values for the color correction factor can be recovered for
all values of $\alpha$ used in this paper from the formula
\BEL{equ_dirbe_k}
K(\alpha, T) = \frac{\sum_n a_n(\alpha) \tau^n}{\sum_m b_m(\alpha) \tau^m},~~
\tau \equiv \log_{10}T,
\EE
where the $a$ and $b$ coefficients may be found in Table
\ref{table_dirbe_k}.

The choice of an $\alpha=2$ emissivity model was not well-motivated
in \cite{sfd98}.  The dust column map is only very weakly dependent
upon the emissivity law because the entire map is renormalized using
direct observations of reddening.
Using an $\alpha=1$ emissivity model changes the relative column density
of dust between warm and cold regions by only $\sim 1\%$.
However, the extrapolated emission at lower frequencies is highly
dependent upon the emissivity of the dust, and the $\alpha=2$
assumption must be tested. 

\subsection{Comparison with FIRAS Data at 500 GHz}
\label{sec_compfiras}

The extrapolated, millimeter emission from dust as predicted from
\cite{sfd98} can be compared directly to the FIRAS measurements.
As a test of this model, we
examine the spatial correlation of FIRAS $500 \GHz$ emission with the
DIRBE $100\micron$ map.  Such a comparison is most meaningful at a
FIRAS frequency that is nearly on the Rayleigh-Jeans tail of the dust
spectrum, but is
still easily measured against the 2.73K CMBR. This strikes a
balance between the poor $S/N$ ratio at lower frequencies and the
stronger temperature-dependence of the spectrum at higher
frequencies. 

To increase S/N, we synthesize a broadband FIRAS $500 \GHz$ channel.
We sum the FIRAS measurements in the 10 unmasked channels, $i$, between
$400$ and $600 \GHz$, weighting by $\nu^{-3.5}$ to make the summand
roughly constant at these frequencies:
\BEL{equ_firas500}
    {\rm FIRAS~500} =
    { \sum_i{ F_i \nu_i^{-3.5} } \over { \sum_i{ \nu_i^{-3.5} } } }
\EE
The \cite{sfd98} extrapolation (equation \ref{equ_extrap1})
is similarly summed over the same FIRAS frequency bins to generate
a predicted broadband flux:
\BEL{equ_extrap2}
    {\rm SFD~500} = { \sum_i{
      K^{-1}_{100}(\alpha,T) I_{100}
      {\nu_i^\alpha B_{\nu_i}(T) \over \nu_0^\alpha B_{\nu_0}(T) }
      \nu_i^{-3.5} }
    \over \sum_i{ \nu_i^{-3.5} } }
\EE
The correlation between FIRAS~500 and SFD~500 is very good
(Figure \ref{fig_firas_scatter}{\it c}), with an RMS dispersion of $0.2 \MJypSr$
about the best-fit line.  For comparison, we plot the correlation
with (a) \HI\ column density, and (b) DIRBE $100\micron$ flux.
Both of these show a scatter that is 3.7 times worse than with SFD~500,
demonstrating that submillimeter emission from dust is neither simply related
to the \HI\ column density, nor is the dust at one temperature everywhere
on the sky.

The SFD98 extrapolations work impressively well in predicting
$500\GHz$ emission from dust, despite their assumptions of $\nu^2$
emissivity and one temperature along each line-of-sight.
However, the slope of the regression between FIRAS~500 and our
extrapolations differs significantly from unity (formally by nearly 
$40\sigma$).
At lower frequencies, the slope departs even more strongly from unity.
This is an indication that a $\nu^2$ emissivity is incorrect for the dust,
as was seen from the mean spectrum of large regions of the sky
(Figure \ref{fig_3temp}).
This will be addressed in detail in \S \ref{sec_spectrum}.

\subsection{The Spectrum of Dust-Correlated Emission}
\label{sec_spectrum}

The many frequency channels of the FIRAS experiment allow detailed
comparisons with predictions for the spectrum of dust emission.
For each channel of the FIRAS data, we compute a correlation slope
with the SFD prediction.  The correlation slope is computed as the
best-fit slope of the FIRAS column ($F_p$) versus the predicted column
($I_p$).  By subtracting a weighted mean from each map, the
correlation slope, $m$, is insensitive to zero-point uncertainties in
either map:
\BE m = \frac{ \langle W_p (F_p - \bar{F_p}) (I_p - \bar{I_p})\rangle}
        { \langle W_p (I_p - \bar{I_p})^2 \rangle }
\EE
where $W_p$ is the FIRAS pixel weight ($W_p\sim 1/\sigma_p^2$) for pixel
$p$.  Such a slope is computed for each FIRAS channel centered at
$\nu_i$.  These slopes are equivalent to that computed in Figure
\ref{fig_firas_scatter}{\it c} for a broad-band FIRAS channel.
The correlation between the FIRAS column and the SFD prediction is strong
and apparently free from systematic errors
in all but the lowest frequency channels.
For all FIRAS channels $\nu < 2100 \GHz$,
we compute $m$ using 71\% of the sky described in \S \ref{sec_mask}.
If the $\nu^2$ emissivity model used by SFD98 were valid,
the slope would be consistent with unity.

\begin{figure}
\plotonesmall{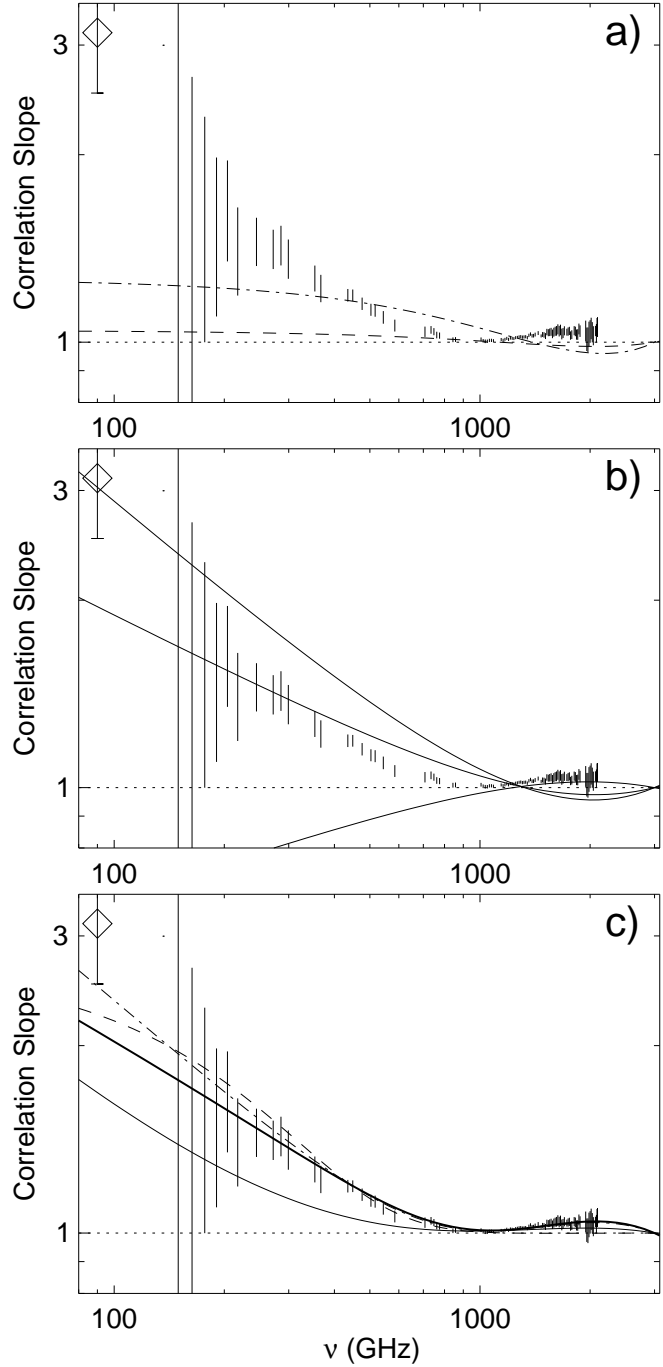}
\caption{Dust-correlated emission, scaled by
$\nu^2B_\nu(\bar{T})$ for ease of comparison.
The FIRAS data (error bars) would be consistent with unity if the
$\nu^2$ emissivity model were correct.
Panel ({\it a}) overplots broadened temperature models with $\Delta
T_{FWHM} = 3\K$
(dashed line) and $\Delta T_{FWHM} = 6\K$ (dash-dot line).
Panel ({\it b}) overplots single-component models with
$\nu^{1.5}$ (top), $\nu^{1.7}$, and $\nu^{2.2}$ emssivity laws.
The horizontal dotted line corresponds to $\nu^{2}$. 
Panel ({\it c}) overplots two-component models, with the best-fit model
shown as a solid line.  See Table \ref{table_results} for the specific
model parameters.
These results are \emph{not} sensitive to an isotropic background in the
FIRAS data.
The DMR $90\GHz$ measurement is shown as a diamond.
The DMR $30$ and $53\GHz$ measurements fall well above any model curves.
}
\label{fig_firas_slope_spec}
\end{figure}

In each panel of Figure \ref{fig_firas_slope_spec}, the correlation
slopes, or ``slope spectrum,'' are plotted as a function of frequency,
with the vertical lines extending to $\pm3\sigma$.  Overplotted in
each panel are various models, evaluated for a typical high-latitude
$\Rmap$ value of 0.68.  To facilitate comparison with the data, these
models are each divided by the same $\nu^2$ prediction as the data.
Such a comparison is instructive, but not used for formal fits,
because the ratio of a given model prediction to the $\nu^2$
prediction depends weakly upon $\Rmap$, and therefore varies across
the sky.  Note that all model ratios are constrained via our
temperature correction to be unity at 100 and $240\micron$ (3000 and
$1250\GHz$).  The data points are also constrained to go through unity
at $1250\GHz$ to the extent that the FIRAS and DIRBE data are consistent.

At $\nu > 500 \GHz$, this slope spectrum is consistent
with the $\nu^2$ model to within $10\%$.
At lower frequencies, the slope spectrum increases, demonstrating that
there is more emission at low frequencies than a $\nu^2$ emission model
would predict.

\subsection{Other Power-Law Emissivities}

There is no power-law emissivity model that fits the FIRAS data.
The SFD prediction can be made with other emissivity profiles by
modifying the exponent, $\alpha$, in equation \ref{equ_extrap1}.
An $\alpha=1.5$ emissivity profile results in a better fit
at low frequencies, but ruins the fit at high frequencies
(see Figure \ref{fig_firas_slope_spec}{\it b}).
An $\alpha=2.2$ emissivity gives a good fit at high frequency, but is
catastrophically wrong at low frequency. 
The minimum $\chi^2$ is achieved for an $\alpha\approx2.0$ extrapolation.
The value of $\chi^2$ doubles for $\alpha=1.5$ (see Table \ref{table_results}).

\subsection{Broadened Temperature Distribution}
\label{sec_broad_temp}

Our model ignores the possibility of dust temperature variation along
a line-of-sight through the Galaxy.  Such a situation may arise
from the superposition of different environments with different temperatures.
Or it may arise from an intrinsic distribution in dust grain sizes and
temperatures within a given environment.
 
The far-IR/submillimeter dust emission is expected to be dominated by
large grains ($0.01\micron < a < 0.25\micron$), 
which are in equilibrium with the ambient radiation
field.  The grains are thought to follow a power law distribution of
grain sizes $dn/da \propto a^{-3.5}$ (\cite{mathis77}) from about
$0.0005\micron$ (the size at which absorption of a single photon can
sublimate mass away from the grain) to $0.25\micron$, where the number
density appears to fall off based on $R_V$ measurements (\cite{kim94}).
Since the long wavelength emissivity of a grain scales as its size $a$
times the surface area, or volume $a^3$, the larger grains dominate
the submillimeter emission (\cite{draine85}).
This will be true unless the slope of the size
distribution is steepened to a slope of nearly $-4.0$.  The very small
grains (VSGs; $a \la 0.01\micron$) are transiently heated and emit at high
effective temperatures for a small fraction of the time, but do not
contribute significantly to submillimeter emission. 

Even for the grains which are large enough to be in equilibrium with
the ambient radiation field, there is a slight size-dependent
temperature variation.  The approximation that the grains are small
compared to the wavelength of absorbed radiation is not exactly
satisfied at the large end of the grain size distribution, so the
larger grains are a bit colder because they absorb less efficiently
relative to their emission.  For reasonable assumptions about the
ISRF, the temperature varies approximately as $a^{-0.06}$ (\cite{draine84}),
both for silicate and graphite grains.  Over the size range of
interest, $0.01\mu < a < 0.25\mu$, the temperature range of the grains
at a given locale is modest.  The dominance of the largest grains' emission
results in a narrow distribution of relevant grain temperatures, and
allows us to use the emission-weighted mean temperature for each
component.  This approximation is good at $100\micron$ and exact on
the Rayleigh-Jeans tail where $I_\nu \propto T$.  This greatly
simplifies our analysis.

There are two other reasons why the temperature might vary along a
line of sight.  The cutoff of the grain size distribution at large $a$ might
vary, causing the dominant size (and temperature) component to vary along
the line of sight.  Another possibility is that the interstellar
radiation field may vary.  Along lines of sight passing through cold
molecular clouds, both of these effects should contribute.  For
extinction predictions (as in SFD98) it is important to understand
which of these effects is causing the temperature variation; for
extrapolating the emission to microwaves, the cause is unimportant. 

In order to model such variations,
we experimented with Gaussian-broadened distributions of temperatures
with width $\Delta T_{FWHM}$ along a single line of sight. 
The ratio of a broadened-$T$ model ($\Delta T_{FWHM}=3\K$)
to the idealized single-temperature fit is shown as the dashed line in Figure
\ref{fig_firas_slope_spec}{\it a}.
An even broader distribution ($\Delta T_{FWHM}=6\K$) is plotted as a dash-dot line.
The broadened-$T$ model changes the predictions by at most $4\%$
for $\Delta T_{FWHM}=3\K$ and 1$5\%$ for $\Delta T_{FWHM}=6\K$.  
These models retain large and coherent deviations from
the FIRAS data.  The models are slightly more consistent with the
low-frequency FIRAS channels, but less consistent with the high-frequency
channels.  The value of $\chi^2$ for these models is higher than that
of the single-temperature $\nu^2$ model.
We conclude that a Gaussian-broadened temperature distribution does not
fit the data any better than a single-temperature model and would only
introduce poorly-constrained parameters into our models.
The lack of an acceptable one-component model -- even when temperature
variation along each line of sight is included -- indicates the need
for multi-component models, which are discussed in the next section. 
 
\section{MULTI-COMPONENT DUST MODELS}
\label{sec_multicomp}

In this section we explain the theoretical motivation for
multi-component dust models, present our general model, and then
provide the results for eight specific models. 

\subsection{Theoretical Motivation}
The diffuse ISM is known to contain many different types of molecules
and dust grains with a broad range of physical properties.  In spite of
the expected melange of dust grains, it was originally expected that
in the far-IR/submillimeter bands, all dust would have similar optical
properties.  For example, Draine \& Lee (1984) predicted $\nu^2$
emissivity for both silicate and graphite grains.  

The emission mechanism corresponding to fundamental vibrations (single
photon/phonon interactions) in crystalline dielectric materials is
optically inactive due to wavevector conservation, and multi-photon
interactions are rare at low temperatures.  Therefore, the emissivity
of crystalline materials would be expected to be dominated by absorption in
the damping wing of an infrared active fundamental vibration, the
strength of which goes as $\nu^2$ at low frequencies.  In a metallic
or semimetallic material, interaction with electrons was expected to
dominate FIR absorption, also resulting in $\nu^2$ emissivity
(\cite{wooten72}).

In amorphous materials, the lack of long range order causes a
breakdown of the selection rules that forbid single photon/phonon
interactions, and all modes become active.  The emissivity power law
then only depends upon the density of states, which was also thought
to go as $\nu^2$ (\cite{kittel76}).  Thus, amorphous materials were
expected to have the same dependence on frequency as other components,
but for an entirely different reason.  The most notable exception to
the $\nu^2$ theory was the case of planar structures such as graphite,
which would yield a $\nu^1$ power law by the same reasoning.  For an
excellent summary of the theoretical details, see \cite{tielens87}.

More recent laboratory measurements suggest that universality of
$\nu^2$ emissivity is an oversimplification, with different species of
grains having differing emissivity laws.  The composition and
abundance of grains of different species can be constrained by
astronomical observations and by observations of solar system bodies.
A multi-component model for interstellar dust has been constructed by
Pollack \etal\ (1994), based on laboratory measurements,
observations of molecular cloud cores, and fits to dust shells with
temperatures $T \approx 100 \K$ around young stars. Their model
predicts that at frequencies $\nu
\simgt 500 \GHz$, dust emission will obey a $\nu^{2.6}$ emissivity law
due to the dominance of carbon species.  At lower frequencies, the
emission is dominated by astronomical silicates such as olivine
([Mg,Fe]$_2$SiO$_4$) and orthopyroxene ([Mg,Fe]SiO$_3$).  This
low-frequency prominence of silicates flattens the emissivity profile
to $\nu^{1.5}$ at frequencies $\nu \simlt 500 \GHz$ ($\lambda \simgt
600\micron$)

Despite the complexities in the dust composition, most authors have
chosen for simplicity to model the observed emission with a single power-law
emissivity.  If one's observations are limited to less
than a decade in frequency, this parameterization may be adequate to fit
the data, especially if one component dominates the emission.
However, combining data from all three \COBE\ instruments results
in a tremendous range in observed frequencies.  
The discussion in \S\ref{sec_1comp}
demonstrates that a single power-law emissivity
is a poor fit to this combined data.  Our physical interpretation is
that different grain species dominate the emission at different
frequencies.

Laboratory measurements of submillimeter-wave absorption properties of
both crystalline and amorphous silicates (\cite{agladze96}) suggest
that $\alpha$ ranges from approximately $1.2 - 2.7$, with some
components having a much higher opacity than others (by a factor of
$\sim 40$ at 300 GHz and 20K).  These studies motivate a broader
search of parameter space.

\subsection{General Multi-Component Model}

We outline a general formalism for describing a mixed population of
dust grains.  These simple considerations apply only in the limit
of large grains ($ a > 0.01\micron$) which are not transiently heated
but instead reach equilibrium with the local radiation field.  We
neglect emission from very small, transiently heated grains because it
is unimportant over the FIRAS frequency range.

\subsubsection{Statement of Model}
Fortunately, for wavelengths $\lambda > 100\mu$ the large grains
totally dominate the thermal (vibrational) dust emission.
Since it is common to assume that each
component of the dust will have a power-law emissivity over the
FIRAS frequency range (\cite{pollack94}), one can sum these to construct
a multi-component model analogous to equation \ref{equ_extrap1}:
\BEL{equ_multi}
   I_{p,\nu} =
   { \sum_k{ f_k Q_k(\nu) B_\nu(T_{pk})}
    \over \sum_k{ f_k Q_k(\nu_0) B_{\nu_0}(T_{pk}) 
	K_{100}(\alpha_k,T_{pk})} } I_{p,100}
\EE
where $f_k$ is a normalization factor for the $k$-th component, $T_{pk}$ is
the temperature in pixel $p$ of component $k$, $K_{100}$ is the DIRBE
color-correction factor (\cite{dirbesupp95}) and $I_{p,100}$ is the SFD98
$100\micron$ flux at pixel $p$ in the DIRBE filter. 
The emission efficiency $Q(\nu)$ is the ratio of the emission cross
section to the classical cross section of the grain.  Because the
grains of interest are very small compared to the wavelength of
emission, $Q(\nu) \propto a$, where $a$ is the radius of a spherical
grain.  One interpretation of this is that the grain is so small that
all parts of the grain are close enough to the grain surface to take
part in the emission. 

The emission opacity (effective area per mass) for a spherical grain
of radius $a$, $\kappa^{em}(\nu)$, is related to $Q(\nu)$ by
\BE
\kappa^{em}(\nu) = \frac{\pi a^2}{\rho V} Q(\nu) =
\frac{3Q(\nu)}{4\rho a}.
\EE
Because $Q/a$ is usually taken to be independent of $a$ for $a<<\lambda$
(\cf, \cite{hildebrand83}), $\kappa^{em}(\nu)$ does not depend on
grain size.  The frequency dependence is taken to be a power law
\BEL{equ_emissivity}
\kappa^{em}(\nu) = \kappa^{em}(\nu_0)(\nu/\nu_0)^{\alpha_k}
\EE
where $\alpha_k$ is the emissivity index and $\kappa^{em}(\nu_0)$ is
the opacity of species $k$ at a reference frequency $\nu_0$ = 3000 GHz.
It will be convenient to interpret $f_k$ as the fraction of power
absorbed and re-emitted by component $k$, so we force the
power fractions to sum to unity:
\BEL{equ_fsum}
  \sum_k f_k = 1 .
\EE

\subsubsection{Temperature Coupling}
The degrees of freedom in this multi-component model can be
substantially reduced by demanding that the components are in
equilibrium with the interstellar radiation field (ISRF).  
If we assume that the ISRF has a constant spectrum everywhere on the
sky and varies only in intensity, then we may define 
$\kapstar_k$ to be the effective absorption opacity (cross-section per
mass) to the ISRF in the limit
of low optical depth:
\BE
  \kapstar_k = { \int_0^{\infty} \kappa^{abs}_k(\nu) I_{ISRF}(\nu) d\nu
  \over \int_0^{\infty} I_{ISRF}(\nu) d\nu }
\EE
where $I_{ISRF}(\nu)$ represents the angle-averaged intensity
in the ISRF as a function of
frequency, and has the same dimensions as $B_\nu$.  To avoid confusion,
$\kappa^{abs}_k(\nu)$ designates the optical opacity of component $k$,
which is physically related to the submillimeter opacity $\kappa^{em}_k(\nu)$
but need not be an extension of the power law expression for
$\kappa^{em}_k(\nu)$ in equation \ref{equ_emissivity}.
The total power absorbed per mass for species $k$ is given by
\BE
  U_k^{in} = \kapstar_k \int_0^{\infty}{ I_{ISRF}(\nu) d\nu } .
\EE
The power is primarily radiated in the far-infrared, and thus is only
sensitive to the far-IR emissivity law.  The power per mass emitted is
\BE
  U_k^{out} = \int_0^\infty{ \kappa^{em}_k(\nu) B_\nu(T_k)d\nu } .
\EE
Demanding that each grain species is in equilibrium with the ISRF
(\eg, $U_k^{in} =  U_k^{out}$),
the energies of all species are related via:
\BEL{equ_balance}
   {1\over\kapstar_i}\int_0^{\infty} \kappa^{em}_i(\nu) B_\nu(T_i)d\nu =
   {1\over\kapstar_j}\int_0^{\infty} \kappa^{em}_j(\nu) B_\nu(T_j)d\nu .
\EE
Using our parameterization of the emissivities (equation \ref{equ_emissivity}),
we can solve for the temperature of one component as a function of the other:
\BEL{equ_Tcouple}
   T_i^{~4+\alpha_i}
   = {q_j Z(\alpha_j) \over q_i Z(\alpha_i)}
   \left({ h\nu_0 \over k_B }\right)^{\alpha_i-\alpha_j} T_j^{~4+\alpha_j}
\EE
where
\BEL{equ_qdef}
   q = \kappa^{em}(\nu_0)/\kapstar
\EE
is essentially the ratio of far-IR emission cross section 
to the UV/optical absorption cross section, and the
integrals are absorbed into the analytic function
\BE
   Z(\alpha) \equiv \int_0^\infty \frac{x^{3+\alpha}}{e^x-1} dx =
   \zeta(4+\alpha)\Gamma(4+\alpha) .
\EE
Henceforth, we shall use only the ratio of opacities, $q_k$,
assuming that the dust temperature is sensitive only to this ratio of
emission to absorption cross sections.  This is not strictly true,
because $\kapstar$ is weakly dependent upon grain size.  However, as
we showed in \S \ref{sec_broad_temp}, the assumption of a single
temperature for each component -- and therefore a single $q_k$ -- in each
locale is justified. 

\subsubsection{Interpretation of $f$, $q$}
Each dust component is therefore described by three global parameters
($f_k$, $q_k$, $\alpha_k$) and one parameter
that varies with position on the sky, $T_k(\vec{x}$).
Because equation \ref{equ_Tcouple} couples the temperature of each component,
there is only one independent temperature (\eg, $T_2$) per line-of-sight.
The interpretation of the $q_k$ as IR/optical opacity ratios is
obvious, but the meaning of the $f_k$ normalization factors is less
clear.  To understand what the $f_k$ are, let us consider the ratio of
the power, $P_i/P_j$, absorbed and re-emitted by components $i$ and $j$:
\BE
\frac{P_i}{P_j} = \frac{f_i}{f_j}
\frac{\int_0^{\infty} q_i (\nu/\nu_0)^{\alpha_i} B_\nu(T_i)d\nu}
     {\int_0^{\infty} q_j (\nu/\nu_0)^{\alpha_j} B_\nu(T_j)d\nu}
\EE
Combining equations \ref{equ_emissivity}, \ref{equ_balance} and
\ref{equ_qdef}, we see that the integrals are equal:
\BE
 P_i/P_j = f_i/f_j .
\EE
Because the $f_k$ sum to unity (equation \ref{equ_fsum}), we identify
$f_k$ with the fraction of power absorbed from the ISRF and emitted in
the FIR by component $k$.  Note that $P_i/P_j$ is independent of
frequency.  Note also that $f_k/\kapstar_k$ is proportional to the
mass fraction.  Therefore, if the optical opacities of all species
were equal (which is unlikely), then $f_i/f_j$ would measure the mass
ratios between species $i$ and $j$.

Whether or not the actual components of the dust physically correspond to these
components, equation \ref{equ_multi} can be thought of as a phenomenological
``expansion set'' for describing the composite dust spectrum.

\subsection{Fit Results for Specific Models}

We now describe our fits to eight different models of the form
described above.  All results from these fits are described in Table
\ref{table_results}.   In the first four we consider only a single
component ($f_1=1$) whose temperature varies on the sky.
We strongly emphasize that none of our models force a constant temperature
everywhere on the sky, which is an extremely poor description of the
data (as can be seen in Figure \ref{fig_firas_scatter}{\it b}).
For a given a dust model, the temperature is
uniquely constrained by the DIRBE $100\micron / 240\micron$ flux ratio
along each line of sight.  We perform single-component fits for
$\alpha=1.5, 1.7, 2.0,$ and $2.2$, obtaining the best reduced $\chi^2_\nu$
at $\alpha=2.0$.  This is in agreement with previous fits to the
FIRAS data (\cite{boulanger96}) and also with earlier theoretical
prejudice (\cite{draine84}).

These results are encouraging, but statistically a $\chi^2=3801$ for 123
degrees of freedom (for the $\nu^2$ model) is completely unacceptable.  The
first two-component model we consider is one designed to replicate
the spectrum in Pollack \etal\ (1994).
The prescription in their paper for their best-fit broken
power law with $\Tmeanone = \Tmeantwo$ 
corresponds in our model to $\alpha_1=1.5, \alpha_2 = 2.6,
f_1=0.25, $ and $q_1/q_2=0.61$.  This results in a considerably better
fit of $\chi^2_\nu= 15.3$ without fitting any new free parameters.
The choices for $\alpha_1$, $\alpha_2$, $f_1$ and $q_1/q_2$ are
based upon other empirical evidence completely independent of the
DIRBE and FIRAS data sets.
The ratio of the Pollack \etal\ model to the strawman $\nu^2$ model is
shown in Figure \ref{fig_firas_slope_spec}{\it c} as a light solid line.
Between $800$ and $1800 \GHz$, where the FIRAS signal is very good, the
model matches the data to approximately 1\% everywhere.
At lower frequencies, the largest deviation is $25\%$.

Allowing $f_1$ and $q_1/q_2$ to float with fixed $\alpha_1, \alpha_2$
provides even better fits.  We attempt to reproduce the
results found in Reach \etal\ (1995), where a component of very cold dust
was proposed to explain the low-frequency excess.  Reach \etal\ used
$\alpha=2$ for both the warm and cold components as a mathematical
convenience. 
Letting our model float with
$\alpha_1=\alpha_2=2$ we obtain $f_1=0.00261$ and $q_1/q_2=2480$.
This model achieves a better fit than any other model tested in the literature
do date, yielding a $\chi^2=1241$, or 10.3/DOF.
A physical interpretation of this combination of parameters in the
context of our models would imply that there is a component constituting
0.26\% of the dust emission power, but with an opacity ratio 2480 times
higher than the dominant component.
This huge opacity ratio explains the low temperature,
$\Tmeanone=5\K$, as compared to the dominant component at $\Tmeantwo=18\K$.
Since this compelling fit appeared in the Reach \etal\ (1995) paper, some
authors have sought to explain the model in the context of this simple
interpretation.
Fractal grains (\cite{fogel98}) and other possibilities have been raised
to explain the opacity ratio, but no convincing mechanism
has yet been proposed.  Opacities may indeed differ
considerably, but factors of many thousand are probably unreasonable.
However, the idea of multiple well-mixed components at different
temperatures deserves further exploration.

By taking the values from Pollack \etal\ (1994) of $\alpha_1=1.5,
\alpha_2=2.6$, but realizing that such different components are very
likely to be at different temperatures, we allow $f_1$ and $q_1/q_2$
to float, obtaining $\chi^2=244$ or 2.0/DOF for $f_1=0.0309$ and
$q_1/q_2=11.2$.  It may seem surprising at first that one component is
11 times ``better'' at thermally radiating than another, but to
justify this we appeal to the empirical measurements of Agladze \etal\
(1996).  They find that the amorphous silicate MgO$\cdot$2SiO$_2$ at
$20\K$, for
example, radiates $\sim 40$ times more readily at 300 GHz than
the crystalline silicate enstatite (MgSiO$_3$) (\cite{agladze96},
Table 1).  The effective optical opacities $\kapstar$ may also vary
significantly, so a wide range in emissivity ratios $q_2/q_1$ is
empirically well established.  Furthermore, our model only requires a
tiny fraction of the dust to be of this kind.  This is a very
reasonable theoretical step to take to obtain a formal increase in
likelihood of $\sim 350$ orders of magnitude over a simple $\nu^2$
model.

A further reduction of $\chi^2=219$ or 1.85/DOF is achieved by
allowing the power law indices $\alpha_1$ and $\alpha_2$ to vary.  The
best-fit values, $\alpha_1=1.67, \alpha_2=2.70, f_1=0.0363,$ and
$q_1/q_2=13.0$.

For these model parameters, the temperatures of the two components
are related by
\BE
   T_1 = 0.352\ T_2^{~1.18}.
\EE
The mean temperatures are $\Tmeanone = 9.4$ and $\Tmeantwo = 16.2$ for
the 71\% of sky that we fit.  This is the model we adopt for the
comparisons discussed in the next section. 

\section{DISCUSSION}
\label{sec_discussion}
\subsection{Interpretation of Best-Fit Model}
The thermal emission from Galactic dust can be very successfully
predicted at millimeter/microwave frequencies using a two-component
composition model with temperature varying on the sky.  We tentatively
refer to the two components as an amorphous silicate-like component
($\nu^{1.7}$
emissivity, $\Tmean \approx 9.5\K$) and a carbonaceous component
($\nu^{2.7}$ emissivity, $\Tmean \approx 16\K$).  
This solution agrees with the FIRAS data much better than a
2-component model using two $\nu^2$ emissivity components with one of
the components very cold ($\Tmean \approx 5\K$) 
Note also that we have obtained only 4 global model parameters from the entire
FIRAS data set - all other column density and temperature information
is derived from the DIRBE $100\micron$ and $240\micron$ maps. 
In the Reach \etal\ analysis, temperatures of the two components were
allowed to float independently, and were fit directly to the FIRAS
data. 

Although our analysis does nothing to rule it out, we find no evidence
for a recently proposed warm component ($\nu^1$ emissivity, $\Tmean
\approx 29\K$) associated with the WIM.  This component results from 
a different approach to modeling the dust emission spectrum (see
Lagache \etal\ 1999 for details).

\subsection{Spatial Coherence of Dust Properties}

Previous two-component models of dust emissivity fit to the FIRAS data
found that the two components must be spatially correlated to a high
degree (\cite{reach95}).  We demonstrate this fact by computing the
correlation slope of dust with FIRAS $500 \GHz$ as shown in Figure
\ref{fig_firas_scatter}.  Removing this correlated emission reduces
the variance in the $500 \GHz$ map by 95\%.  Based on the data at high
Galactic latitude, about 4\% of the variance is attributable to
measurement noise.  This leaves $1\%$ of the variance (or 10\% of
the signal) as the upper limit for uncorrelated $500 \GHz$ emission.  If
there exists a separate cold dust component which does not emit at 
$100\micron$, then it must be very highly spatially correlated with warm dust. 

\subsection{Evidence for Variations in Dust Properties}

\begin{figure}
\plotone{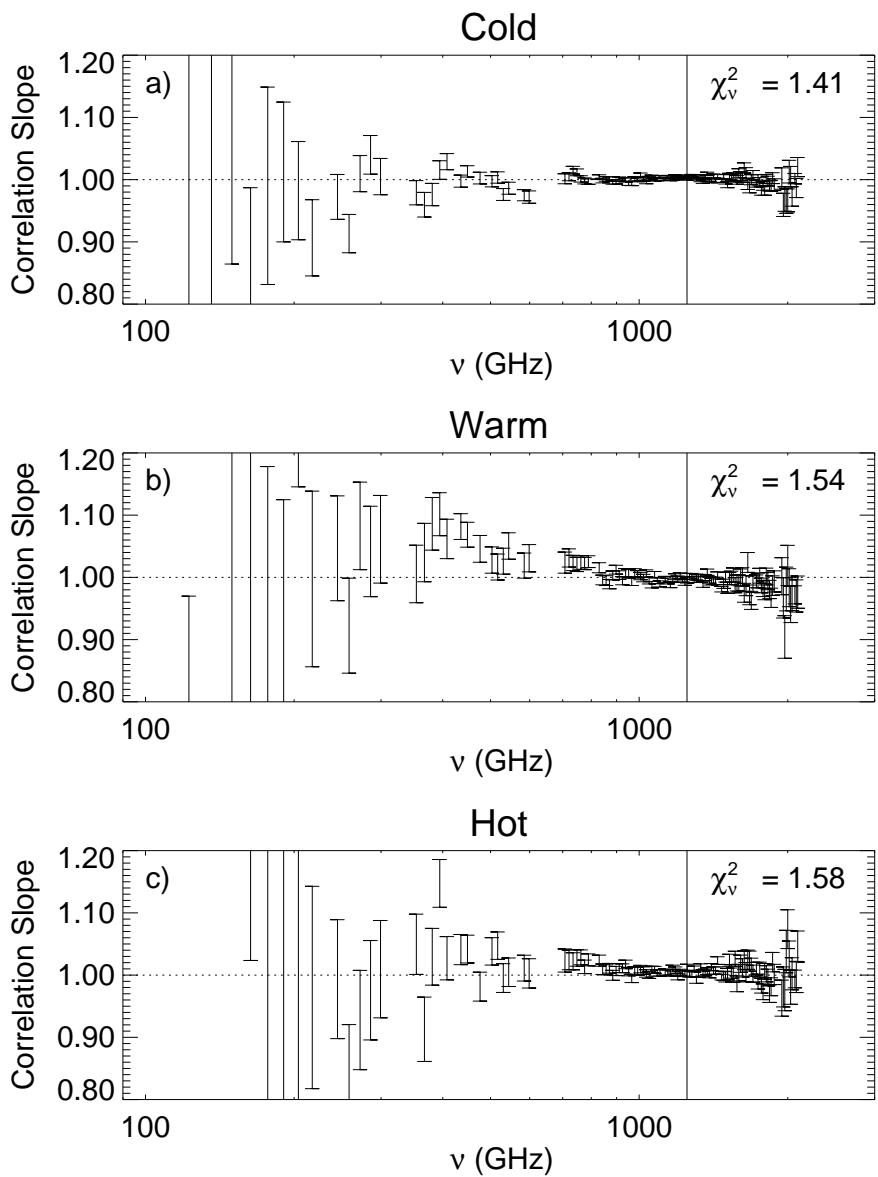}
\caption{FIRAS versus best-fit model correlation slopes.
The sky is divided into three zones based upon temperature:
({\it a}) cold regions ($\Rmap = I_{100\micron}/I_{240\micron} < 0.62$),
({\it b}) warm regions ($0.62 < \Rmap < 0.69$), and
({\it c}) hot regions ($\Rmap > 0.69$).
The systematic residuals between zones is not more than $\sim 5\%$.
The vertical line is drawn at $240\micron$, where the models are constrained
to fit the DIRBE measurements.
}
\label{fig_fitres_temp}
\end{figure}

\begin{figure}
\plotone{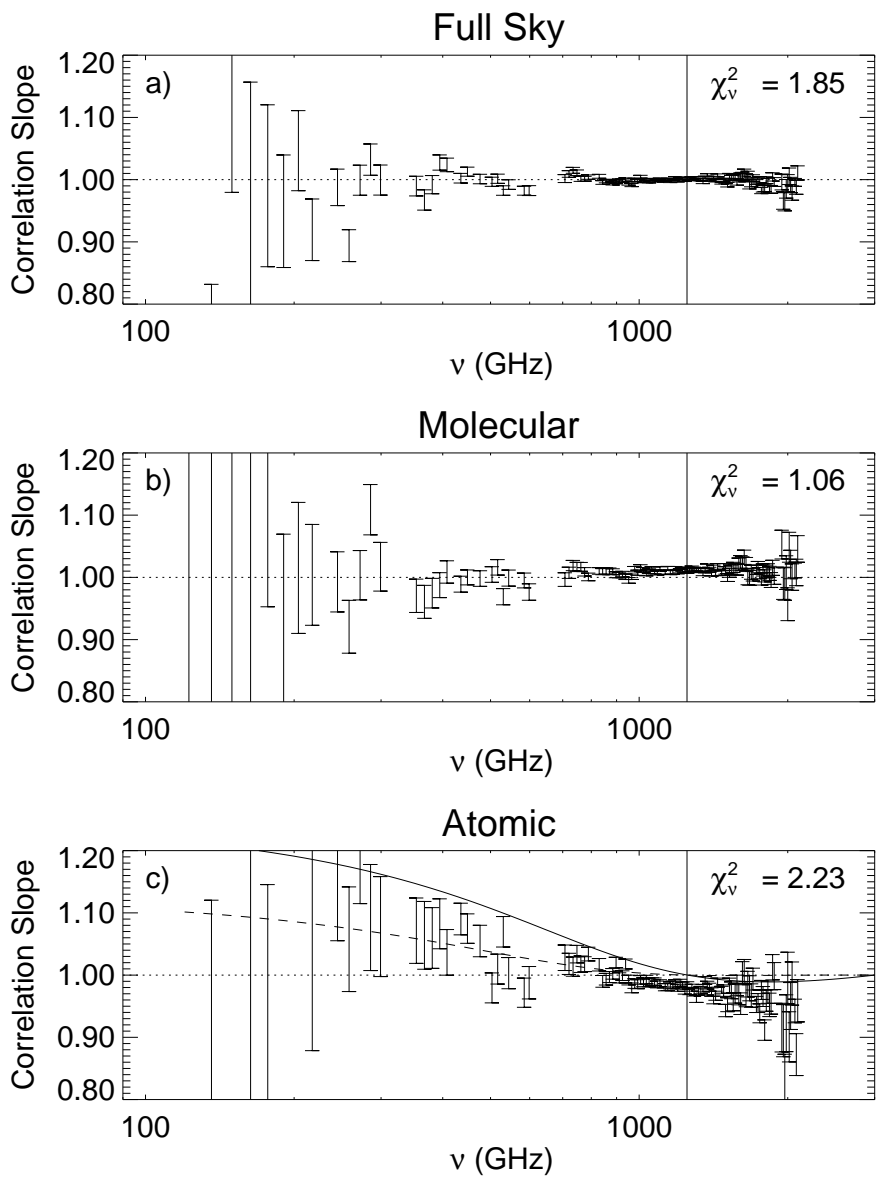}
\caption{FIRAS versus best-fit model correlation slopes.
The sky is divided into two zones based upon dust/gas ratio:
({\it a}) both atomic- and molecular-dominated zones,
({\it b}) zones dominated by molecular clouds, and
({\it c}) zones dominated by atomic gas.
The maximum deviation is $\sim 15\%$ at $100 \GHz$ for the atomic zone.
For the atomic gas, we overplot our best-fit model modified to $f_1=0.0465$
(solid line) or modified to $q_1/q_2=15.0$ (dashed line).
The vertical line is drawn at $240\micron$, where the models are constrained
to fit the DIRBE measurements.
}
\label{fig_fitres_env}
\end{figure}

The agreement between the best-fit two-component model and the FIRAS
data is impressive (see Figures \ref{fig_fitres_temp} and
\ref{fig_fitres_env}).  The reduced $\chi_\nu^2$ is close to unity,
implying that the model uncertainties are small compared to the
measurement errors.  The far-IR/millimeter sky at $|b| \ga 15\degree$
appears to be well-fit by a fixed model for the interstellar dust.
The temperature varies, but the composition and size distribution of
the dust grains are constrained to be very similar everywhere in the
diffuse ISM.

Although our best-fit model appears to successfully describe the
average dust emission spectrum, there still might be systematic
variations across the sky.  Splitting the sky into different zones
based upon various observables, one can search for regions that deviate from
our model.  Dividing the sky according to temperature or dust column
density, we do not find significant differences (see Figure
\ref{fig_fitres_temp}).

However, splitting the sky according to dust/gas ratio, one does find
coherent differences.  We construct a dust/gas ratio by dividing the
SFD98 dust map by the Leiden-Dwingeloo \HI\ map (\cite{dwingeloo}),
both smoothed to a $45\arcmin$ FWHM Gaussian beam.  The SFD98 dust map
is proportional to the $100\micron$ emission expected if the entire
sky were at a uniform temperature.  Regions where the
dust/gas ratio exceeds the high-latitude average
by more than a factor of 2 are designated ``molecular'' (14\% of the
sky).  Remaining pixels are designated ``atomic''(40\%).  
The remaining 46\% of the sky is excluded by our FIRAS mask, or by the
lack of Leiden-Dwingeloo survey data at $\delta<-30\degree$.

In Figure \ref{fig_fitres_env}, we plot the correlation slope between
our model predictions and the FIRAS data for the ``molecular''
and ``atomic'' sky.  Since the full-sky fits are dominated by
the ``molecular'' sky, the model very nearly fits that zone with
a correlation slope near unity at all frequencies
(Figure \ref{fig_fitres_env}{\it b}).  However, ``the atomic'' sky shows
deviations relative to the model that approach $\sim 15\%$ at low
frequencies (Figure \ref{fig_fitres_env}{\it c}).

We suspect that these differences between ``atomic'' and
``molecular'' zones represent true variations in dust grain
properties.  These variations can be quantified by adjusting our model
parameters to achieve a better fit in the ``atomic'' zone.
The fits can be improved by retaining the best-fit parameters in the
molecular zone, and adjusting $\alpha_1$ lower, $f_1$ higher,
or $q_1/q_2$ higher in the atomic zone.
Lowering $\alpha_1$ does not improve the fit as well, and would require
a qualitative change in the millimeter opacities of the ``silicate'' component.
We consider this possibility the least likely, though it is possible
that we are seeing ice mantle accretion or some other environment-dependent
mechanism.

\begin{figure}
\plotonesmall{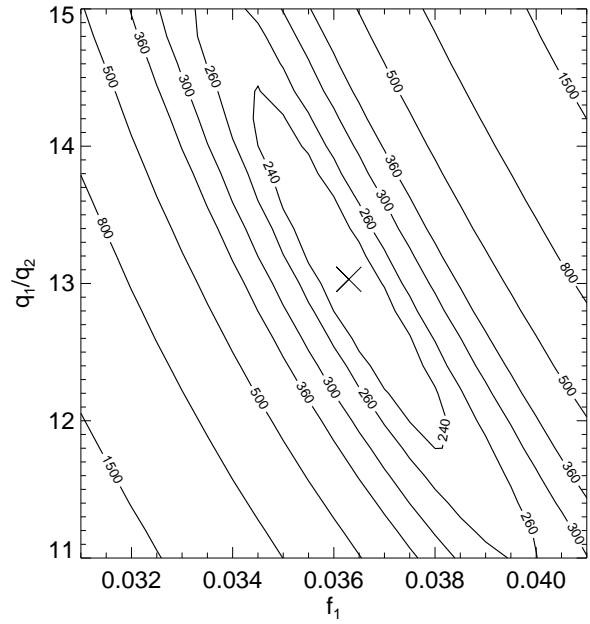}
\caption{Contour of $\chi^2$ for parameters $f_1$ and $q_1/q_2$,
fixing the emissivity laws ($\alpha_1$, $\alpha_2$) to their best-fit values.
Our best-fit two-component model is denoted by an X.
}
\label{fig_fq_con}
\end{figure}

In this region of parameter space, $f_1$ and $q_1/q_2$ are sufficiently
degenerate that an almost equally good fit may be obtained in the atomic
zone by adjusting either parameter a modest amount (Figure
\ref{fig_fq_con}).  Either $f_1$ can
be increased by $\sim 15\%$ or $q_1/q_2$ increased by $\sim 25\%$
in the atomic zone (see Figure \ref{fig_fitres_env}{\it c}).
Increasing $f_1$ is equivalent to increasing the amount of the ``silicate''
component relative to the warmer ``carbon-based'' component.
The change in $q_1/q_2$ could very plausibly be interpreted as a decrease
in $\kapstar_2$ (an increase in $q_2$) in the molecular zone.  If the
second component is physically composed of carbon-based grains, it might
be responsible for the UV absorption bump at $2175\Ang$ and sensitive
to saturation of that feature.  Much of the molecular zone has an
\cite{sfd98} extinction $\Av \ga 0.5\MAG$ which corresponds to
$A(2175\Ang) \ga 1\MAG$.  When this feature begins to saturate, the
carbon grains see a change in the ISRF that effectively reduces $\kapstar_2$.
The small differences we see between the ``atomic'' and ``molecular''
zones are certainly consistent with spatial variations in the
radiation field, which we consider to be one of the most reasonable
explanations.  Note that we have not proved that this mechanism is
responsible, or even that the two dominant components are silicate and
carbon-dominated grains.  However the data are consistent with this
interpretation. 

\section{APPLICATION TO CMBR ANISOTROPY}
\label{sec_dmr}

We compare predictions from our best-fit two-component model to the
\COBE\ DMR data.  Significant microwave emission from dust was found
by Kogut \etal\ 1996 by using DIRBE $140\micron$ flux as a dust
template.  As a similar comparison, we compute the correlation slope
in each DMR channel with respect to our model.  This slope is
sensitive only to emission correlated with the dust, and does not
depend on isotropic backgrounds.  For the purposes of these fits, DMR
pixel $i$ is weighted by the inverse of $\sigma_i^{~2}+\sigma_{CMB}^{~2}$
where $\sigma_i$ is the measurement noise in pixel $i$ and
$\sigma_{CMB}$ is the RMS power in the CMB anisotropy, taken to be
$30\mu$K (\cite{bennett96}). 

\begin{figure}
\plotone{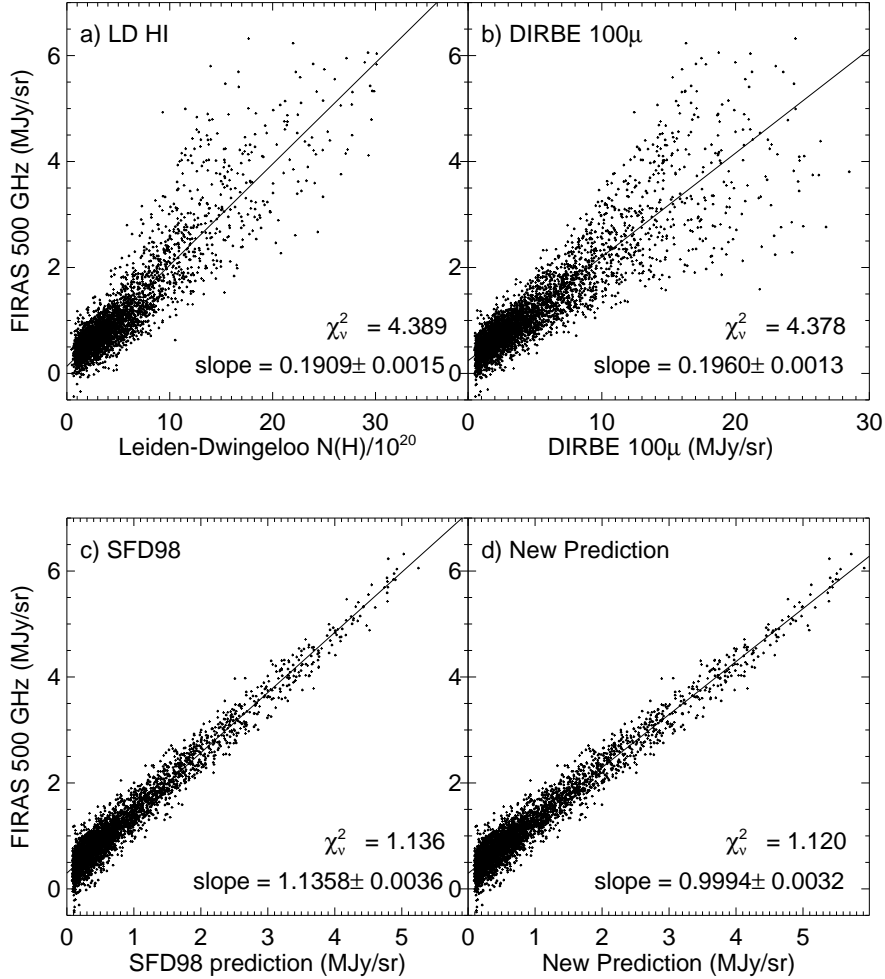}
\caption{FIRAS - DIRBE comparison.
Comparison of FIRAS emission in a synthesized $500\GHz$ broad-band versus
({\it a}) \HI\ emission,
({\it b}) DIRBE $100\micron$ ($3000\GHz$) emission with zodiacal
contamination removed,
({\it c}) prediction from SFD98 using single-component, $\nu^2$-emissivity
model, and
({\it d}) prediction from our best-fit two-component model.
The comparisons are made over 71\% of the sky.
Straight lines are fit and overplotted using the statistical errors in
the FIRAS data.  The scatter about this line is $\sim 3.5$ times
smaller in ({\it c}) or ({\it d}) as compared to ({\it a}) or ({\it b}).
The slope in ({\it d}) is almost unity, as expected for a good prediction.
}
\label{fig_firas_scatter}
\end{figure}

The first column in Table \ref{table_dmr} is the correlation slope of
DMR against our best-fit model evaluated at $500\GHz$.  This frequency
is chosen because it is low enough to be on the Rayleigh-Jeans side of
the Planck function, so the dust spectral index between DMR and our
500 GHz predictions has very little temperature dependence.  Also,
this frequency is high enough that FIRAS obtained high quality data
for the dust emission.
As Figure \ref{fig_firas_scatter} has demonstrated, our model is well
tested at $500\GHz$ and one may be confident that it represents real
dust emission on the sky.  Therefore, $500\GHz$ is a sensible
reference frequency to use in such comparisons.  The next column of
the table is the RMS power in $\mu$K brightness temperature implied by
this correlation slope.  Note that we are confined to the mask used in
the computation of the DMR CMBR anisotropy.  A less exclusive mask
would yield a higher RMS power.  The remaining columns of Table
\ref{table_dmr} are similar, but use the $140\micron$ DIRBE map instead
of our prediction, for direct comparison with the Kogut \etal\
results.  However, as we have already seen in Figure
\ref{fig_firas_scatter}, the correlation at $500\GHz$ is not tight,
and is probably worse at DMR frequencies.  These comparisons in Table
\ref{table_dmr} are meant to indicate the spectral shape of
dust-correlated microwave emission.

It is clear in Table \ref{table_dmr} that the dust-correlated emission
at 31 GHz is larger than at 53 GHZ - when according to our model it
should have fallen by a factor of $\sim 5$.  Comparison of DMR data
with our model evaluated at the same frequency gives an idea of the
amount of the excess.  Again, correlation slopes are computed, for DMR
data as a function of model predictions, with a slope of unity
corresponding to an accurate model.  The results are tabulated in
Table \ref{table_dmr_x}.  At $90\GHz$, there is 20\% more
dust-correlated emission than predicted.  At $53$ and $31.5\GHz$, this
emission is too high by a factor of $2.4$ and $20$, respectively.
Similar excess emission is seen at $14.5$ and $31\GHz$ from the data
of Leitch \etal\ (1997) and at similar frequencies in the Saskatoon
data (\cite{sk94}; \cite{sk95};
\cite{doc97}).

This excess microwave emission is clearly correlated with the dust,
but not due to its thermal (vibrational) emission.  It has recently
been suggested that magnetic dipole emission from paramagnetic grains
(\cite{draine99}) or electric dipole emission from rapidly spinning
dust grains (\cite{draine98b}) could dominate at these frequencies.
Others have suggested that dust-correlated free-free emission may be
responsible (\cite{doc98a}).  However, Draine and Lazarian argue
against this on energetic grounds in \cite{draine98a}.  Galactic
synchrotron emission is not a favored explanation because it is
unlikely to be highly correlated with the dust.

\subsection{Templates for CMBR Contamination}

It is critical that CMBR experimentalists compare their observations
with valid models for the Galactic dust emission.  A ``template approach''
is often carelessly used to compare observations with expected contaminants,
with the correlation amplitude indicating the level of contamination.
For example, $100\micron$ emission maps (\eg, \IRAS\ or DIRBE) or
$21\cm$ maps (\eg, Leiden-Dwingeloo: \cite{dwingeloo})
are often used as templates for microwave dust emission.
These templates ignore well-measured variations in dust temperature
and variations in the dust/gas ratio.
We demonstrate this point by differencing the broad-band FIRAS $500\GHz$
map (Figure \ref{plate_fir500}) with best-fit templates.
The residuals with respect to the \HI\ template (Figure \ref{plate_res_hi})
or the $100\micron$ template (Figure \ref{plate_res_100}) are noticeably
worse than with our model prediction (Figure \ref{plate_res_pred}).
In addition, the \HI\ or $100\micron$ template offers no insight as to
frequency-dependence of the dust emission.  Because 100mu is so
sensitive to the temperature, it may be the worst of these at high
Galactic latitude. 

This need to use the proper template will grow with future data sets.
In the case of the DMR data, no adverse effects resulted from the use
of the $140\micron$ template, as can be seen in Table
\ref{table_dmr}.  The expected RMS power from dust in the DMR channels
is not significantly altered by using our model.  However, the
signal-to-noise ratio of DMR is much less than 1 per pixel.  The
measurement noise overwhelms the template errors in this case.  In the
case of S/N$\sim 1$ data, \eg\ FIRAS 500 GHz, it is apparent from
Figure \ref{fig_firas_scatter} that our template is more
accurate, and much more readily detected, than $100\micron$ flux or \HI\
column.  These considerations will be even more important with
satellites such as \MAP\ ($13'$) because our map is the only full-sky
well calibrated dust model at high resolution ($6'$).  

We would encourage CMBR researchers to present measurements of
dust-correlated microwave emission by using our predictions as a
baseline, so that (at least where vibrational dust emission dominates)
the comparison is free of temperature-dependent biases and assumptions
about the dust/\HI\ ratio.  This will allow easy comparison of
samples from various parts of the sky - a comparison which is quite
difficult with current dust templates. 


\section{SUMMARY AND CONCLUSIONS}
\label{sec_summary}

We have demonstrated that the SFD98 $100\micron$ emission map,
extrapolated with a very simple two-component dust model, is an
excellent predictor of the Galactic emission as seen by FIRAS at all
frequencies.  Although the older SFD98 $\nu^2$ emission predictions
are tightly correlated with the FIRAS data, the correlation slope
deviates significantly from unity at frequencies far from $1250 \GHz$
and $3000 \GHz$ ($240$ and $100\micron$) where it is constrained by
the DIRBE data.  The $\nu^2$ emissivity assumed by SFD98 produces a
reduced $\chi_\nu^2 \approx 30$ when compared with the FIRAS data.  Although
this fit is unacceptable, no other single-component power law
emissivity model improves $\chi_\nu^2$ significantly.

We provide a general multi-component model where each component is
described by an emissivity power law, $\alpha$, power fraction, $f$, and
a ratio of thermal emission to optical opacity, $q$.  Each component
is required to be in equilibrium with the ISRF.  This couples the
temperatures of each component, both of which are constrained along
each line-of-sight by the DIRBE $100/240\micron$ ratio. 

In addition to one-component models evaluated for various emissivity
power law indices, we evaluate 4 two-component models.  They
correspond to Pollack \etal, Reach \etal, our best fit using Pollack
\etal\ emissivities, and our best overall fit for all four model
parameters.  Our best fit parameters are physically reasonable, given
empirical evidence found in Agladze \etal\ (1996).  See Table
\ref{table_results} for a summary of our results.

The data argue very strongly that the dust properties of the ISM are
uniform over virtually the entire high latitude sky on angular scales
greater than $7\degree$.  We have found marginal evidence for
variations in the molecular-dominated zones relative to
atomic-dominated zones.  We tentatively suggest that these variations
are due to UV optical depth effects within the molecular zones.

This thermal (vibrational) dust emission model fails to explain
dust-correlated microwave emission observed by DMR.  The $90\GHz$
emission is in approximate agreement with our model, but the $53$ and
$31\GHz$ DMR emission is high by factors of $2.4$ and $20$,
respectively.  This excess emission could result from rapidly spinning
dust grains (\cite{draine98b}) or from free-free emission.  Whatever
the emission mechanism, it must be strongly correlated with the
thermal (vibrational) dust emission.

Predictions of our best-fit model for thermal dust emission will be
made available on the World Wide Web.\footnote{The data are publicly
available via the World Wide Web at
{http://astro.berkeley.edu/dust}.}


\section{ACKNOWLEDGMENTS}

Conversations with Dale Fixsen contributed enormously to our
understanding of the FIRAS data.  We would also like to thank Bruce
Draine, Carl Heiles, Dave Hollenbach, Chris McKee, David Spergel, Eric
Gawiser, and Jeffrey Newman for encouragement and helpful discussions.
An anonymous referee helped clarify a number of points. 
Computers were partially provided by a Sun AEGP Grant. DPF is an NSF
Graduate Fellow.  DJS is partially supported by the \MAP\ project and
by the Sloan Digital Sky Survey.  This work was supported in part by
NASA grants NAG 5-1360 and NAG 5-7833.  The \COBE\ datasets were
developed by the NASA Goddard Space Flight Center under the guidance
of the \COBE\ Science Working Group and were provided by the NSSDC.

\newpage
\appendix
\section{FURTHER PROCESSING OF THE FIRAS DATA}
\label{app_firas}

The FIRAS pass 4 data products are released in two sets of
frequency bins, corresponding to the two sides of the instrument.
There are 43 bins in the low frequency (LOWF) set, running from $68.02$
to $639.37 \GHz$, and 170 bins in the high frequency (HIGH) set, running
from $612.19$ to $2911.29 \GHz$.  The bin spacing is $13.6041 \GHz$, giving a
3 bin overlap between LOWF and HIGH.

\subsection{Recalibration}

An error in the FIRAS external calibrators is described in \cite{mather99}.
The thermometers were found to be miscalibrated by $5\mK$,
causing a systematic error in LOWF.  It is expected that this miscalibration
introduces an error of less than one percent in the LOWF data, which is 
negligible for the purposes of this paper.  Therefore we have chosen to ignore
this problem in the data. 

Comparisons of the high-frequency FIRAS data to the DIRBE $240\micron$
data shows an inconsistency at the $1\%$ level.  The DIRBE gain is
uncertain at this level owing to the uncertainties in its filter
response, and calibration technique.  However, we have chosen to
reduce all the FIRAS measurements by $1\%$, which is well within the
gain uncertainty of the HIGH data, and within the measurement noise of
the LOWF data.  Because the covariance of neighboring FIRAS frequency
bins is embodied in the FIRAS covariance matrix, our results are only
weakly dependent upon this $1\%$ recalibration.

\subsection{Bad Bins}

Line emission from CO, [C~I], [N~II], [C~II], [O~I], and CH was
detected.  We have excluded those bins and bins corresponding to
O$_2$, H$_2$O, and [Si~I], even though no emission was detected by the
FIRAS team.  A few bins were excluded because of residuals in the
mirror transport mechanism (MTM) ghost removal.  Several bins are
excluded from our analysis in the frequency range $639.37$ to $680.21
\GHz$ for three reasons: (1) inefficiencies in the dichroic splitter,
(2) a destructive interference pattern caused by reflection off of the
plastic holder of one of the optical elements, and (3) an aliased MTM
sideband.  All of these effects taken together overwhelm the signal in
these bins and justify their exclusion.  A summary of the frequency
bins excluded in our analyses is found in Table \ref{table_badbins}.

Our analyses make use of 123 frequency bins at $100 < \nu < 2100
\GHz$.  At lower frequencies, the S/N of the dust emission is less
than one.  At higher frequencies, the absolute error in the FIRAS gain
exceeds $2\%$ due to uncertainties in the bolometer calibration (see
\cite{firas_supp}, \S 7.3.2).


\section{COMPUTATIONAL METHODS}
\label{app_compute}

Comparison between the Galactic dust emission as observed by FIRAS and
our predictions is computationally challenging.
The predictions are made at the resolution of DIRBE to take full
advantage of temperature information on scales of $\sim 1\degree$.
These DIRBE-resolution predictions are then smoothed to the FIRAS beam and
compared to 4376 FIRAS pixels ($71\%$ of the sky)
at 123 frequencies, for a total of $\sim 540,000$ data value comparisons.

\subsection{Computation of Spectral Shape}
Aside from the 4 global parameters $(f_1, q_1/q_2, \alpha_1, \alpha_2)$,
the model predictions are made \emph{only} from the DIRBE observations
at $100$ and $240\micron$.  We make use of the DIRBE $100\micron$ map
described in \cite{sfd98} with zodiacal light and the cosmic infrared
background removed.  Our dust temperatures are always derived from 
a ratio map, $\Rmap$, which is a filtered $I_{100}/I_{240}$
flux ratio in the DIRBE passbands.  For a multi-component dust model,
this ratio map measures the following combination of model parameters:
\BEL{equ_Rmap}
  \Rmap_p = { \sum_k{ K_{100}(\alpha_k,T_{pk}) I_{100}(T_{pk}) }
   \over    \sum_k{ K_{240}(\alpha_k,T_{pk}) I_{240}(T_{pk}) } }
\EE
For more than one component, the temperatures are related via
equation \ref{equ_Tcouple}.
For each model, we tabulate $\Rmap$ as a function of the warmer
component, $T_2$, as described by equations \ref{equ_Tcouple} and
\ref{equ_Rmap}.  We fit a 6-th order polynomial to the curve
$\ln T_2(\ln\Rmap)$ for the domain $10 < T_2 < 31\K$.
For our best-fit two-component model,
\begin{eqnarray}
  \ln{T_2} = 2.872
            + 0.2407    \ln{\Rmap}
            + 2.962\times 10^{-2}   \ln^2{\Rmap} 
            + 4.719\times 10^{-3}  \ln^3{\Rmap} \\ \nonumber
            + 9.320\times 10^{-4} \ln^4{\Rmap}
            + 1.109\times 10^{-4} \ln^5{\Rmap} .
\end{eqnarray}
At each DIRBE pixel $p$, we read the values of the $100\micron$ flux
and the ratio $\Rmap$, which in turn recovers $T_1$ and $T_2$.
The flux at any frequency is then given by equation \ref{equ_extrap1}
for a one-component model or equation \ref{equ_multi} for models with
more than one component.

Conceptually, the DIRBE-based predictions are convolved with
the FIRAS beam before comparison to the FIRAS data.
In practice, such a straight-forward approach proved too computationally
expensive when minimizing the residuals over several model parameters.
The dust temperature variations are rarely large within one FIRAS beam,
allowing us to make approximations for the temperature distribution
within a beam.
Let $i$ index the 6144 FIRAS pixels and $j$ index the DIRBE map pixels.
Within each FIRAS beam centered on FIRAS pixel $i$, our predictions
can be explicitly expressed as
\BEL{equ_fmodel}
  F_i(\nu) = \sum_j B_{ij} I_j Y(\nu, \Rmap_j)
\EE
where $B_{ij}$ describes the beam pattern (the fractional contribution
of pixel $j$ to FIRAS pixel $i$), $I_j$ is the $100\micron$ flux
in DIRBE pixel $j$, and $Y(\nu,\Rmap_j)$ describes the model spectral shape,
\BE
  Y(\nu, \Rmap_j) = { I_j(\nu,\Rmap_j) \over I_j(\nu_0,\Rmap_j) } .
\EE
The beam pattern is normalized to unity,
\BE 
  \sum_j B_{ij} = 1 ,
\EE 
for each FIRAS pixel.
Equation \ref{equ_fmodel} can be rewritten as
\BEL{equ_fmodel2}
  F_i(\nu) = \Imeani \sum_j W_{ij} Y(\nu, \Rmap_j)
\EE
where we have defined a weighted mean for the $100\micron$ flux,
\BE
  \Imeani = \sum_j B_{ij} I_j ,
\EE
and a weighting function $W_{ij}$ defined as
\BE
  W_{ij}=\frac{B_{ij} I_j}{\sum_j B_{ij} I_j} 
\EE
This weighting function is also normalized to unity within each beam:
\BE
  \sum_j W_{ij} = 1 .
\EE
A direct evaluation of \ref{equ_fmodel} would work, but is very
expensive, so we resort to a Taylor expansion. 

\subsection{Taylor Expansion}
At each frequency $\nu$, $Y$ depends only upon the $100\micron/240\micron$
flux ratio, $\Rmap$.
We expand $Y(\Rmap)$ about the weighted mean ratio in FIRAS pixel $i$,
$\Rbar$, as follows:
\BE
  Y(\Rmap_j) = Y(\Rbar) + Y'(\Rbar)(\Rmap_j - \Rbar) +
  {1\over 2} Y''(\Rbar) \left[ \Rmap_j - \Rbar \right]^2 +
  {1\over 6} Y'''(\Rbar) \left[ \Rmap_j - \Rbar \right]^3 + \cdots
\EE
where the derivatives are with respect to $\Rmap$ and we have
dropped the $\nu$ subscript for clarity.
Computing the weighted sum of $Y$ within one FIRAS pixel yields
\BEL{equ_Yexpand}
  \sum_j W_{ij} Y(\Rmap_j) = Y(\Rbar) + {1\over 2} Y''(\Rbar) \sigma_i^2(\Rmap)
  + {1\over 6} Y'''(\Rbar) s_i^3(\Rmap)  + \cdots
\EE
where the term linear in $(\Rmap_j - \Rbar)$ vanishes, and $\sigma_i^2(\Rmap)$,
\BE
  \sigma_i^2(\Rmap) = \sum_j W_{ij} (\Rmap_j - \Rbar)^2
\EE
is a weighted variance within FIRAS pixel $i$ and 
\BE
  \s_i^3(\Rmap) = \sum_j W_{ij} (\Rmap_j - \Rbar)^3
\EE
is a weighted difference cubed.  Combining equations \ref{equ_fmodel2}
and \ref{equ_Yexpand} the flux at any frequency is recovered via
\BEL{equ_expand}
  F_i(\nu) = \Imeani
  \left[{ Y(\Rbar) + {1\over 2} Y''(\Rbar) \sigma_i^2(\Rmap)
          + {1\over 6} Y'''(\Rbar) s_i^3(\Rmap) }\right] .
\EE
Note that this expansion is implemented to describe temperature
fluctuations within a FIRAS pixel.
The values of $\Imeani$, $\Rbar$, $\sigma_i^2(\Rmap)$ and $s_i^3(\Rmap)$ 
need only be computed once for all the DIRBE values
within each FIRAS pixel.  Once these values are saved, there is no need
to return to the higher-resolution DIRBE maps.
The predictions for a given dust model establishes the relationship
$Y(\Rmap)$, and the flux in FIRAS pixel $i$ is quickly computed via
equation \ref{equ_expand}.

We have carried the Taylor expansion to third order to establish convergence:
the third-order terms are significantly smaller than the second-order
terms, and are usually negligible.
All results in this paper are obtained with the third-order Taylor series. 
Setting $\sigma^2(\Rmap)$ equal to zero would ignore these small-scale
temperature variations, and introduce errors at the level of a few percent
in our model predictions.

\subsection{Definition of $\chi^2$}
The comparison between predictions and the FIRAS data is further
simplified by collapsing the problem spatially.  At each frequency, a
regression line is computed for the FIRAS flux as a function of the
predicted flux (as in Figure \ref{fig_firas_scatter}).  The pixel
weights from the FIRAS data are used for these regressions.  The
zero-point of the best-fit slope is ignored, as it is sensitive to
uncertainties in the zodiacal light model or the cosmic infrared
background (CIB).  The slope of
the regression is our measure of goodness-of-fit for a model, with a
slope of unity at all frequencies corresponding to perfect agreement.

The $\chi^2$ for each model is computed from the 123 slope values and
significance of their deviations from unity.  The full covariance
matrix $C_{ij}$ (\cite{firas_supp}, \S 7.1.2) for the FIRAS data is used
to couple the errors between frequency bins.
We define a dimensionless covariance,
\BE
  C_{ij}' = \frac{ C_{ij} }{\sqrt{C_{ii} C_{jj}}}
\EE
where $i$, $j$ index the 123 used frequency bins.  The variance in the
correlation slope $m$ at frequency $i$, $\sigma^2(m_i)$, is derived
from the linear
regression for each frequency, assuming uncorrelated Gaussian
measurement noise.  These $\sigma^2(m_i)$ values are dimensionless,
because the $m_i$ are dimensionless slopes of order unity.  Because of
the frequency covariance, the variance at frequency $i$ contains
contributions from the measurement errors at all frequencies $j$ as:
\BE
  \sigma^2(m_i) = \sum_j{ \sigma^2(m_j) C_{ij}' }.
\EE
This covariance matrix does not include the contribution from the
overall bolometer gain errors $J_i$ (termed JCJ errors in \cite{firas_supp},
\S 7.3.2).
The full covariance matrix includes the JCJ terms $J_i J_j$,
yielding a $\chi^2$ of
\BE
\chi^2 = m_i [C'_{ij} \sigma(m_i) \sigma(m_j) + J_i J_j]^{-1} m_j .
\EE
This expression for $\chi^2$ is used for the fits in this paper. 


\section{DATA PRESENTATION}

We provide an electronic data distribution that computes thermal
emission from Galactic dust for any of the models considered in
this paper.  The preferred model is the two-component model with
$\alpha_1=1.67$, $\alpha_2=2.70$.  Intensities are computed at
any frequency using equation \ref{equ_extrap1} for single-component
models and \ref{equ_multi} for two-component models.  The sky brightness
is computed in units of$\MJypSr$ which can be multiplied by
$ 4024(\nu/90\GHz)^{-2} \mu$K$(\MJypSr)^{-1}$ to convert to a brightness
temperature in $\mu$K.  Brightness temperature may be converted to
thermodynamic $\Delta T$ by multiplying by the ``planckcorr'' factor:
\BEL{equ_planckcorr}
{\rm Planckcorr} = \frac{(e^x-1)^2}{x^2 e^x}
\EE
where $x=h\nu/k_bT_{CMB}$ and $T_{CMB}=2.73$
In Table \ref{table_conversion}, these factors are evaluated for a
number of frequencies typical of CMB anisotropy experiments.

The predictions for thermal emission discussed in this paper are
based upon the DIRBE $100\micron$ map (with zodiacal light and CIB
removed) and a ratio map, $\Rmap$, which is a filtered $I_{100}/I_{240}$
flux ratio in the DIRBE passbands.
These maps are stored as simple FITS images in pairs
of $1024\times 1024$ pixel Lambert ZEA (Zenithal Equal Area) polar
projections, similar to the data format used for \cite{sfd98}.
The NGP projection covers the northern Galactic
hemisphere, centered at $b=+90\degree$, with latitude running clockwise.
The SGP projection covers the southern Galactic hemisphere,
centered at $b=-90\degree$, with latitude running counterclockwise.
(Note that Figs.\ \ref{plate_fir500}, \ref{plate_res_hi}, \ref{plate_res_100},
and \ref{plate_res_pred} show the SGP projections rotated by $180\degree$.)
Galactic coordinates ($l,b$) are converted to pixel positions ($x,y$) via
\begin{eqnarray}
  x = {N\over 2} \sqrt{1 - n \sin{(b)}}~\cos{(l)} + {(N-1)\over 2} \\
  y = - {nN\over 2} \sqrt{1 - n \sin{(b)}}~\sin{(l)} + {(N-1)\over 2}
\end{eqnarray}
where $N=1024$ and $n=+1$ for the NGP, and $n=-1$ for the SGP.
Pixel numbers are zero-indexed, with the center of the lower left pixel
having position $(x,y)=(0,0)$.
These Lambert projections are minimally distorted at high Galactic latitudes,
with the distortion approaching $40\%$ at $b=0\degree$.
The pixel area of $(9{\farcm}49)^2$ oversamples the FWHM of $40'$.

Predictions can be made at higher resolution by extrapolating from the
\IRAS\ rather than the DIRBE $100\micron$ map.  We use the high-resolution
$100\micron$ map from \cite{sfd98}, which contains reprocessed \IRAS/ISSA
data recalibrated to DIRBE.
These maps contain $4096\times 4096$ pixels ($N=4096$).
The pixel size of $(2{\farcm}372)^2$ well samples the FWHM of $6{\farcm}1$.
This map has $\sim 20,000$ \IRAS\ sources removed, which is
appropriate for microwave predictions since IR-luminous stars and galaxies
are not expected to contribute significantly to the microwave sky brightness.

The caveats to using these maps to predict emission from Galactic dust
can be summarized as follows:
\begin{enumerate}
\item At frequencies on the Wien tail of the emission
  ($\lambda \simlt 100\micron$),
  we underestimate the dust emission by not including the contribution
  from small, transiently heated grains.
\item  At microwave frequencies ($\nu \simlt 100 \GHz$),
  we have ignored magnetic dipole emission (\cf, \cite{draine99})
  and electric dipole emission from rapidly
  rotating grains (\cf, \cite{draine98b}).  Either of these mechanisms
  may be expected to dominate the thermal emission.
\item Our best-fit model is not a complete description of the dust
  properties.  This model shows residuals that correlate with such
  environmental properties as the dust temperature and dust/gas ratio.
\item Unresolved infrared-luminous Galactic sources (primarily stars)
  are removed from the \IRAS\ maps to a flux level of
  $f_{100} \approx 0.3\Jy$ (see \cite{sfd98}).  These stars are not expected
  to contribute significantly to the sky brightness at frequencies
  $\nu \simlt 1000\GHz$, but this has not been explicitly tested.
\item Although the angular resolution is $40'$ for the DIRBE $100\micron$
  map and $6{\farcm}1$ for the reprocessed \IRAS\ $100\micron$ map,
  our extrapolations to other frequencies relies on an $\Rmap$ map with
  an effective resolution of $1.{\degree}3$.
\end{enumerate}

The data files and corresponding software will be available in the CD-ROM
series of the AAS, or from
our web site.\footnote{http://astro.berkeley.edu/dust/index.html.}
Mask files are also available that contain the most important processing
steps for any given position on the sky.
Further details will be available with the data files.


\bibliographystyle{unsrt}
\bibliography{gsrp}


\clearpage

\begin{figure}
\plotplate{fir500.ps}
\caption{
FIRAS broad-band $500\GHz$ map, as defined in the text
(equation \ref{equ_firas500}).
These Lambert ZEA polar projections are centered on the NGP (left),
and SGP (right),
with Galactic latitude labeled in degrees.  Lines of constant latitude
and longitude are spaced every 30 degrees.
}
\label{plate_fir500}
\end{figure}

------------------------------------------------------------------------------
\begin{figure}
\plotplate{res_hi.ps}
\caption{
Difference map between the broad-band FIRAS $500\GHz$ map and the best-fit
\HI\ template convolved to the same beam shape.
Declinations south of $\delta = - 30\degree$ were not observed in the
Leiden-Dwingeloo (Hartmann \& Burton 1996) survey, accounting for the
blank regions of missing data.  The thin black lines outline FIRAS
pixels masked from our analysis (see \S \ref{sec_mask}).
}
\label{plate_res_hi}
\end{figure}

\begin{figure}
\plotplate{res_100.ps}
\caption{
Difference map between the broad-band FIRAS $500\GHz$ map and a best-fit
$100\micron$ template convolved to the same beam shape.
}
\label{plate_res_100}
\end{figure}

\begin{figure}
\plotplate{res_pred.ps}
\caption{
Difference map between the broad-band FIRAS $500\GHz$ map and the emission
predicted from our best-fit two-component dust model convolved to the same
beam shape.
The residuals are noticeably smaller than residuals obtained by using
either an \HI\ template (Figure \ref{plate_res_hi}) or $100\micron$ template
(Figure \ref{plate_res_100}).
}
\label{plate_res_pred}
\end{figure}


\clearpage
\begin{deluxetable}{c|l|rrrr|rrr|r|r}
\footnotesize
\tablewidth{0pt}
\tablecaption{Fit results for dust emission models
   \label{table_results}
}
\tablehead{
   \colhead{\#}          &
   \colhead{Model}       &
   \colhead{$\alpha_1$}  &
   \colhead{$\alpha_2$}  &
   \colhead{$f_1$}       &
   \colhead{$q_1/q_2$}   &
   \colhead{$\Tmeanone$} &
   \colhead{$\Tmeantwo$} &
   \colhead{$P_1/P_2$}   &
   \colhead{$\chi^2$}    &
   \colhead{$\chi^2_\nu$}
}
\startdata
1 & One-component: $\nu^{1.5}$ emis & 1.5  & -    & 1.0    & 1.0  & 20.0 & -    &  -    & 24943 &  204 \nl
2 & One-component: $\nu^{1.7}$ emis & 1.7  & -    & 1.0    & 1.0  & 19.2 & -    &  -    &  8935 &   73 \nl
3 & One-component: $\nu^{2.0}$ emis & 2.0  & -    & 1.0    & 1.0  & 18.1 & -    &  -    &  3801 &   31 \nl
4 & One-component: $\nu^{2.2}$ emis & 2.2  & -    & 1.0    & 1.0  & 17.4 & -    &  -    &  9587 &   79 \nl
5 & Pollack \etal\ 2-component      & 1.5  & 2.6  & .25    & 0.61 & 17.0 & 17.0 & .33   &  1866 &  15.3 \nl
6 & Two-component: both $\nu^{2}$   & 2.0  & 2.0  & .00261 & 2480 &  4.9 & 18.1 & .0026 &  1241 &  10.3 \nl
7 & Two-component: fit $f,q$        & 1.5  & 2.6  & .0309  & 11.2 &  9.6 & 16.4 & .0319 &   244 &  2.03 \nl
8 & Two-component: fit $f,q,\alpha_1,\alpha_2$					      
                                    & 1.67 & 2.70 & .0363  & 13.0 &  9.4 & 16.2 & .0377 &   219 &  1.85 \nl
\enddata
\tablecomments{The dust models are described by $\alpha_1$, $\alpha_2$,
$f_1$, and $q_1/q_2$.  The mean temperatures for each dust component,
$\Tmeanone$ and $\Tmeantwo$, are evaluated for the
mean $I_{100}/I_{240}$ color ratio in the high-latitude sky.
The ratio of power emitted by each component is $P_1/P_2$.
}
\end{deluxetable}
\clearpage
\begin{deluxetable}{lrrrr}
\footnotesize
\tablewidth{0pt}
\tablecaption{FIRAS bad frequency channels
   \label{table_badbins}
}
\tablehead{
   \colhead{Reason}        &
   \colhead{Channel}       &
   \colhead{$\nu$}         &
   \colhead{$\nu$}         &
   \colhead{$\lambda$}     \\
   \colhead{}              &
   \colhead{}              &
   \colhead{($\cm^{-1}$)}  &
   \colhead{($\GHz$)}      &
   \colhead{($\micron$)}  
}
\startdata
CO (J=1-0)     &      3 &  3.631 &  108.83 &  2753.8 \nl
CO (J=2-1)     &     12 &  7.717 &  231.27 &  1295.9 \nl
MTM            &     18 & 10.440 &  312.90 &   957.8 \nl
MTM            &     19 & 10.894 &  326.50 &   917.9 \nl
CO (J=3-2)     &     20 & 11.348 &  340.10 &   881.2 \nl
O$_2$          &     26 & 14.072 &  421.73 &   710.6 \nl
CO (J=4-3)     &     29 & 15.433 &  462.54 &   647.9 \nl
[CI]           &     31 & 16.341 &  489.75 &   611.9 \nl
H$_2$O         &     36 & 18.611 &  557.77 &   537.3 \nl
CO (J=5-4)     &     37 & 19.065 &  571.37 &   524.5 \nl
Dichroic       &     40 & 20.427 &  612.19 &   489.6 \nl
Dichroic       &     41 & 20.881 &  625.79 &   478.9 \nl
Dichroic       &     42 & 21.335 &  639.40 &   468.7 \nl
\hline
Dichroic       &  43-48 & 21.6   &  646    &   464   \nl
CO (J=6-5)     &     49 & 23.150 &  693.81 &   432.0 \nl
CO (J=7-6)     &     57 & 26.782 &  802.65 &   373.4 \nl
[CI]           &     58 & 27.236 &  816.25 &   367.2 \nl
H$_2$O         &     80 & 37.222 & 1115.54 &   268.7 \nl
[NII]          &    105 & 48.570 & 1455.65 &   205.9 \nl
H$_2$O (J=2-1) &    124 & 57.195 & 1714.12 &   174.8 \nl
[CII]          &    137 & 63.096 & 1890.98 &   158.5 \nl
[CII]          &    138 & 63.550 & 1904.58 &   157.4 \nl
[CII]          &    139 & 64.004 & 1918.19 &   156.2 \nl
[CII]          &    140 & 64.457 & 1931.79 &   155.1 \nl
[OI]           &    149 & 68.543 & 2054.23 &   145.9 \nl
[Si I]         &    168 & 77.167 & 2312.71 &   129.6 \nl
[NII]          &    179 & 82.161 & 2462.35 &   121.7 \nl
CH (J=2-1)     &    188 & 86.246 & 2584.79 &   115.9 \nl
\enddata
\end{deluxetable}

\clearpage
\begin{deluxetable}{l|rr|rr}
\footnotesize
\tablewidth{0pt}
\tablecaption{Comparison of dust templates with DMR
   \label{table_dmr}
}
\tablehead{
   \colhead{$\nu$}              &
   \colhead{Model($500\GHz$)} &
   \colhead{RMS}   &
   \colhead{DMR/$140\micron$}   &
   \colhead{RMS} 
\\
   \colhead{GHz} &
   \colhead{slope$\times 10^{-3}$} &
   \colhead{$\mu$K} &
   \colhead{slope$\times 10^{-4}$} &
   \colhead{$\mu$K} 
}
\startdata
31.5 & $ 1.81 \pm 0.27 $ & $ 20.5 \pm 3.1 $ & $ 1.81 \pm 0.28 $ & $ 20.6 \pm 3.2 $ \nl
53   & $ 1.35 \pm 0.29 $ & $  5.4 \pm 1.1 $ & $ 0.99 \pm 0.29 $ & $  4.0 \pm 1.2 $ \nl 
90   & $ 4.27 \pm 1.08 $ & $  5.9 \pm 1.5 $ & $ 3.38 \pm 1.10 $ & $  4.7 \pm 1.5 $ \nl 
\enddata
\tablecomments{Correlation slopes of the DMR channels with best-fit
500 GHz prediction and $140\micron$ DIRBE data.  Slope values are
dimensionless flux ratios.  RMS values are the RMS power, in $\mu$K,
expected in the DMR maps due to the dust emission traced by template. 
These values compare to an RMS power of $29\pm 1 \mu$K due to CMBR anisotropy.
}
\end{deluxetable}
\clearpage
\begin{deluxetable}{l|rrr}
\footnotesize
\tablewidth{0pt}
\tablecaption{Excess dust-correlated DMR emission
   \label{table_dmr_x}
}
\tablehead{
   \colhead{$\nu$}              &
   \colhead{DMR/Model($\nu$)}     &
   \colhead{$\%$ variance} &
   \colhead{RMS power} 
\\
   \colhead{GHz} &
   \colhead{} &
   \colhead{} &
   \colhead{$\mu$K} 
}
\startdata
31.5 & $ 20.09 \pm 3.09 $  & 0.87 & $  20.2 \pm 3.1 $ \nl
53   & $  2.45 \pm 0.50 $  & 0.57 & $   5.6 \pm 1.1 $ \nl
90   & $  1.18 \pm 0.29 $  & 0.42 & $   6.2 \pm 1.5 $ \nl
\enddata
\tablecomments{Excess dust-correlated microwave emission measured by
DMR.  1) frequency of DMR channel, in GHz.  2) correlation slope of
DMR emission vs. model predictions.  3) percent of variance in DMR
data accounted for by this dust.  Note that the vast majority of the
variance is receiver noise.  4) RMS power due to dust emission, in
$\mu$K brightness temperature.
}
\end{deluxetable}
\clearpage
\begin{deluxetable}{l|rrrrrrr}
\footnotesize
\tablewidth{0pt}
\tablecaption{DIRBE K-correction fit coefficients
   \label{table_dirbe_k}
}
\tablehead{
\colhead{coefficient} &
\colhead{$ \alpha= 1.50$} &
\colhead{$ \alpha= 1.67$} &
\colhead{$ \alpha= 1.70$} &
\colhead{$ \alpha= 2.00$} &
\colhead{$ \alpha= 2.20$} &
\colhead{$ \alpha= 2.60$} &
\colhead{$ \alpha= 2.70$} 
}
\startdata
$a_0$ &  1.00000 &   1.00000 &   1.00000 &   1.00000 &   1.00000 &   1.00000 &   1.00000 \nl
$a_1$ &  2.08243 &   2.15146 &   2.14106 &   2.18053 &   2.55941 &   3.16383 &   3.31600 \nl
$a_2$ & -4.72422 &  -4.84539 &  -4.83639 &  -4.89849 &  -5.41290 &  -6.23131 &  -6.43306 \nl 
$a_3$ &  2.29118 &   2.35210 &   2.35919 &   2.38060 &   2.57867 &   2.86900 &   2.93939 \nl  
$b_0$ & -0.88339 &  -0.87985 &  -0.93625 &  -0.80409 &  -0.80318 &  -0.50356 &  -0.41568 \nl  
$b_1$ &  4.10104 &   4.10909 &   4.19278 &   3.95436 &   4.20361 &   4.07226 &   4.02002 \nl  
$b_2$ & -4.43324 &  -4.43404 &  -4.46069 &  -4.27972 &  -4.55598 &  -4.70080 &  -4.72432 \nl  
$b_3$ &  1.76240 &   1.76591 &   1.77103 &   1.70919 &   1.80207 &   1.87416 &   1.88865 \nl  
\enddata
\tablecomments{Fit coefficients for DIRBE $100\micron$ band color
correction factors $K_{100}(\alpha,T)$, fitted using equation
\ref{equ_dirbe_k}.
}
\end{deluxetable}
\clearpage
\begin{deluxetable}{l|rrrr}
\footnotesize
\tablewidth{0pt}
\tablecaption{Unit Conversion Factors for Selected CMBR Experiments 
   \label{table_conversion}
}
\tablehead{
\colhead{Experiment} &
\colhead{$\nu$} &
\colhead{$\lambda$} &
\colhead{factor} &
\colhead{Planckcorr}
\\
\colhead{} &
\colhead{GHz} &
\colhead{mm} &
\colhead{$\mu$K/(MJy/sr)} &
\colhead{}
}
\startdata
COBE/DMR      &    31.5 &    9.52 &   32849    &     1.02582 \nl
              &    53.0 &    5.66 &   11603    &     1.07448 \nl
              &    90.0 &    3.33 &    4024    &     1.22684 \nl \hline
MAP           &    22.0 &   13.64 &   67344    &     1.01253 \nl
              &    30.0 &   10.00 &   36216    &     1.02340 \nl
              &    40.0 &    7.50 &   20371    &     1.04190 \nl
              &    60.0 &    5.00 &    9054    &     1.09623 \nl
              &    90.0 &    3.33 &    4024    &     1.22684 \nl \hline
Planck        &    30.0 &   10.00 &   36216    &     1.02340 \nl
              &    44.0 &    6.82 &   16836    &     1.05087 \nl
              &    70.0 &    4.29 &    6652    &     1.13275 \nl
              &   100.0 &    3.00 &    3259    &     1.28562 \nl
              &   143.0 &    2.10 &    1594    &     1.65110 \nl
              &   217.0 &    1.38 &     692.2  &     2.98186 \nl
              &   353.0 &    0.85 &     261.6  &    12.8186  \nl
              &   545.0 &    0.55 &     109.7  &   157.85    \nl
              &   857.0 &    0.35 &      44.38 & 15392       \nl \hline
\enddata
\tablecomments{
Column 1 contains the frequency, in GHz, for which the unit conversion
factors are
computed.  Column 2 is the corresponding wavelength in mm.  A value in
units of MJy/sr should be multiplied by the factor in Column 3 to
convert to $\mu$K brightness temperature.  Brightness
temperature is multiplied by Planckcorr (Column 4; equation
\ref{equ_planckcorr}) to convert to
thermodynamic temperature, assuming $T_{CMB}=2.73$K.}
\end{deluxetable}
\clearpage

\end{document}